\documentclass[twocolumn]{aastex7}

\usepackage{array}
\usepackage{amsmath,amstext}
\usepackage[T1]{fontenc}
\usepackage{natbib}
\usepackage{sidecap}
\usepackage{enumitem}
\usepackage{appendix}
\usepackage{caption}
\usepackage{subcaption}
\usepackage{float}
\usepackage{accents}
\usepackage{booktabs}

\begin{document}

\begin{flushright}
        \includegraphics[width=6cm]{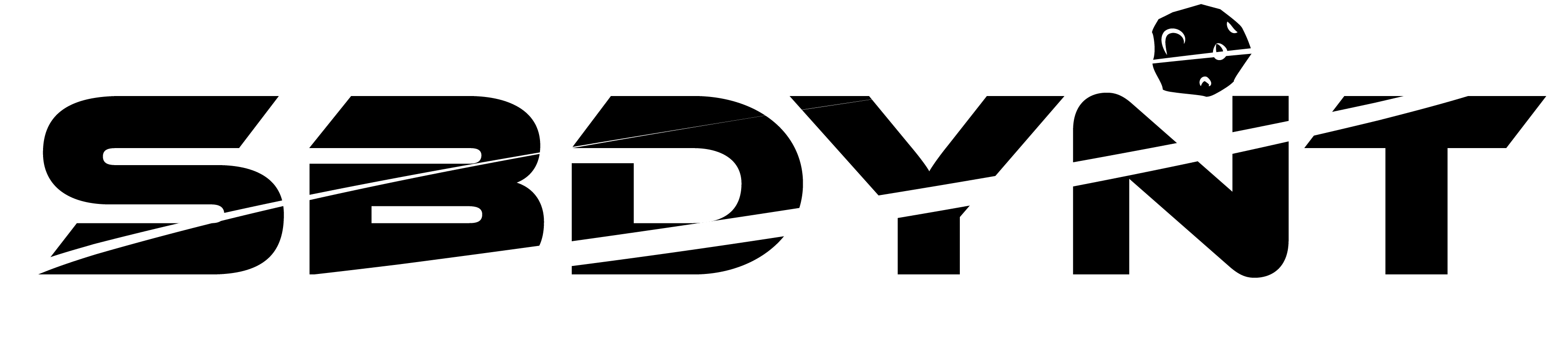} % adjust width as needed
    \end{flushright}
    
    \vspace*{-0.5cm} % Adjust vertical space from top
    
\title{On the Secular Evolution of Inclinations Among the Classical Belt TNOs} 

\author[0000-0003-4051-2003]{Dallin Spencer}
\affiliation{Brigham Young University, Department of Physics and Astronomy, N283 ESC, Provo, UT 84602, USA}
\email{djspenc@byu.edu}

\author[0000-0003-1080-9770]{Darin Ragozzine}
\affiliation{Brigham Young University, Department of Physics and Astronomy, N283 ESC, Provo, UT 84602, USA}
\email{darin_ragozzine@byu.edu}

\author[0000-0001-8736-236X]{Kat Volk}
\affiliation{Planetary Science Institute, 1700 E Fort Lowell Rd STE 106, Tucson, AZ 85719, USA}
\email{kvolk@psi.edu}

\author[0000-0002-1226-3305]{Renu Malhotra}
\affiliation{University of Arizona, Lunar and Planetary Laboratory, 1629 E University Blvd, Tucson, AZ 85721, USA}
\email{malhotra@arizona.edu}

\shorttitle{}

%%%%%%%%%%%%%%%%%%%%%%%%%%%%%%%%%%%%%%%%%%%
\begin{abstract}

Proper orbital elements have been used in diverse ways to study the dynamical architecture of transneptunian objects (TNOs), providing insights into Solar System evolution and migration theories.
In this paper, we present the numerically-computed proper elements of all known TNOs.
We then compare their distribution with the distribution of proper elements of a large synthetic sample generated to study the secular behavior of TNOs in the Classical belt region in an effort to better understand certain observed features, particularly features appearing in the inclination distribution of these TNOs. 
Our analysis indicates the presence of an inclination-dependent instability in the Classical belt from $42<a<45$ au, which we find is caused by the $g+s-g_8-s_8\approx0$ secular resonance. 
The coincidence of this instability with the previously-noted inclination boundary of the Cold Classical belt  near proper inclination of $I_{free} = 4^{\circ}$ could suggest that the primordial Classical belt may have originally been more continuous between $4^{\circ} < I_{free} < 6^{\circ}$ than we observe today. 
However, our analysis to account for the impact of secular resonant instabilities finds a best-estimate of the inclination width of the primordial Cold Classical belt near $\approx1.5^{\circ}$, well below $4^\circ$, and slightly thinner than previous works have suggested.
We conclude that the inclination boundary of the Cold Classical belt is not set directly by secular resonance or instability, and that this belt is generally unaltered by resonance or instability since the end of planetary migration. 
\end{abstract}
%%%%%%%%%%%%%%%%%%%%%%%%%%%%%%%%%%%%%%%%%%%


%%%%%%%%%%%%%%%%%%%%%%%%%%%%%%%%%%%%%%%%%%%
%%%%%%%%%%%%%%%%%%%%%%%%%%%%%%%%%%%%%%%%%%%
\section{Introduction}\label{s:intro}

Solar System small bodies (hereafter SSSBs) have long contributed important observational constraints on models of solar system formation and evolution.
An early example includes \cite{Hirayama:1918}, who identified clusters in asteroid orbital elements, providing evidence for the very first asteroid collisional families.
These types of studies have been enhanced by the use of proper orbital elements, which more clearly describe the orbital structure of the small bodies as they would be without the forced motion of the giant planets on their orbits \citep[e.g.][]{Milani:1994, Knezevic:2003, Spencer:2026}.
In particular, the dynamical architecture of the Trans-Neptunian objects (TNOs) exists as a primordial relic of the early Solar System; as such, the study of TNO proper elements can provide great insight into theories of early planetesimal formation as well as the signatures of planetary migration. \citep[e.g.]{Knezevic:2003, Brown:2004, Dawson:2012, Morbidelli:2014, VanLaerhoven:2019, Huang:2022}.

A particularly well-studied region of Trans-Neptunian space is  the Classical belt, which contains the largest observed population of small bodies beyond Neptune, extending from $40 < a < 48$ au.
The Classical belt is further divided into separate dynamical classes, the low-inclination Cold Classical TNOs (CCTNOs), which are generally believed to have formed in situ \citep[see, e.g.][]{Morbidelli:2020, Gladman:2021}, and the higher-inclination Hot Classical TNOs (HCTNOs).
The cold and hot populations also display strong divergence beyond simple inclination. 
Compared to the HCTNOs, the CCTNOs densely populate their orbital region, have distinctively red colors \citep[e.g.][]{Pike:2017,Fraser:2023,Bernardinelli:2025}, have a higher binary fraction \citep[e.g.][]{Noll:2020}, and have a deficit of large objects \citep[e.g.][]{Kavelaars:2021,Petit:2023}, though recent analyses indicate that the size distribution of the CCTNOs and HCTNOs may in be a single consistent distribution \citep{Kavelaars:2021,Bernardinelli:2025}.
The physical properties of TNOs are generally expected to relate directly to their formation circumstances (and thus the formation of the solar system as a whole), so the strong correlation between physical properties and current heliocentric orbits in the Classical belt motivates further study into these populations.

For example, recent analyses modeling the streaming instability of planetesimal formation show that the primordial planetesimal disk would display some specific dynamical features which would be preserved in a dynamically cold region.
\cite{Nesvorny:2019} indicates that the streaming instability would cause in situ formation of small body binaries, with a specific rate of mutual inclinations between objects.
\cite{Kavelaars:2021} further found that the streaming instability model predicts a specific size distribution of TNOs, which matches the observations reported by the OSSOS survey in 2021. 
Both of these analyses rely heavily on properly identifying the Cold Classical belt TNOs separately from the Hot Classical TNOs, with the assumption that the CCTNOs have been largely dynamically untouched since their formation, and thus represent an immediately post-formation population of objects.  

The recent development of the Small-Body Dynamics Tool (SBDynT) software provides an efficient and easy method for quickly performing robust dynamical analysis of Solar System small bodies through the computation of stability indicators and synthetic proper orbital elements by any user \citep{Spencer:2026}.
Previous catalogs of TNO proper elements exist \citep[e.g. AstDys-2]{Knezevic:2000}, but often don't report proper or mean elements for objects which lie near resonance, or experience larger forced terms.
In addition, the previous catalogs are updated infrequently, motivating the development of a more efficient and user-friendly tool for the computation of new proper elements for the growing catalog of small solar system bodies.
Here we present a newly computed catalog of TNO proper elements for all known TNOs, with a focus on better understanding the secular dynamics of the Classical belt (TNOs with semimajor axes in the range $a=40-48$~au). 
Then, to better understand the distribution of Classical belt TNO proper elements with respect to the rest of the Solar System, we use SBDynT to measure the proper precession frequency distribution of the Classical belt TNOs, which informs the long-term orbital behavior of these small bodies.

We investigate the relationship between the secular dynamics and proper elements of the Classical belt TNOs, with specific emphasis given to better understanding the lower-inclination Classical belt in the following steps.

\begin{enumerate}
    \item We compute the proper and mean elements for the entire catalog of known TNOs (Section~\ref{sec:prop_catalog}).
    \item We compute proper elements of a large sample of synthetic TNO particles in a dense grid in the Classical belt region to measure their secular behavior and proper frequency distribution throughout proper element space (Section~\ref{sec:sec_mapping}).
    \item We highlight a particular inclination dependent instability associated with a secular resonance which coincides with the location of the previously empirically-defined Cold Classical belt boundary near $I_{free}\approx 4^{\circ}$. 
    We explain the dynamical mechanisms behind this secular resonance and the implications for TNO dynamical evolution (Section~\ref{sec:secular_architecture}). 
    \item We estimate the inclination width of the ancient Cold Classical belt using a new method which accounts for the effect of slow diffusion and instability in inclination (Section~\ref{sec:fitting_results}).
    
\end{enumerate}

Our results provide new dynamical insight into the previous empirically-determined separation between the Cold Classical and Hot Classical TNO populations \citep[e.g.][]{Brown:2001, Brown:2004, VanLaerhoven:2019, Huang:2022}, and provide critical justification for the present-day shape of the Cold Classical belt in inclination space.

\section{TNO Proper Elements Catalog}
\label{sec:prop_catalog}

The most recently published catalog of proper elements can be found in the AstDys-2 catalog.\footnote{See \url{https://newton.spacedys.com/astdys2/index.php?pc=5}}
However, this catalog only contains the computed proper elements for stable TNOs and only for TNOs known at the time (currently reported as 06/2024). 
In addition to proper elements for stable TNOs, our new catalog includes the computation of the ``mean'' elements for less stable and/or resonant objects, which still represent a helpful filtering of the most significant forced terms in particle orbits and still provide critical insight into the average motion of these particles.

\begin{figure*}[!ht]
    \centering
    \includegraphics[width=1\linewidth]{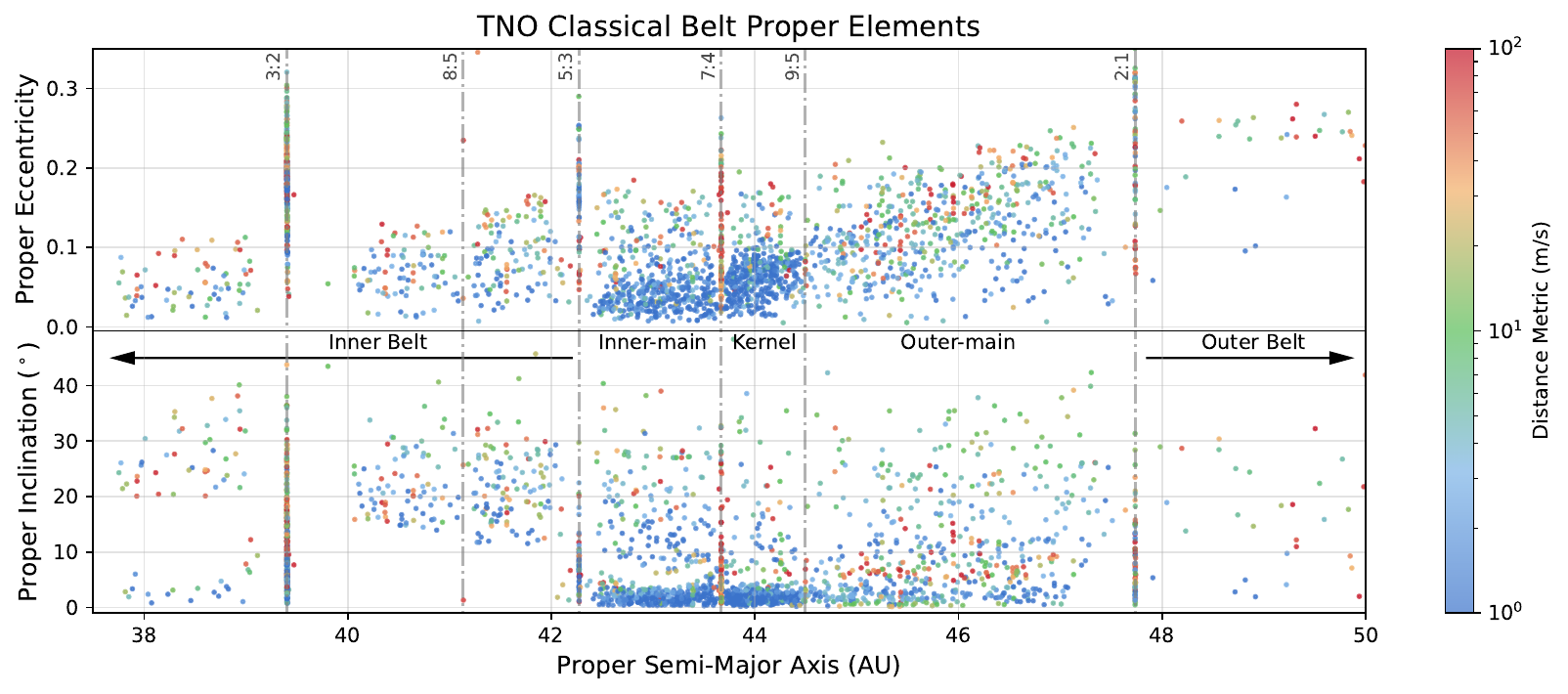}
    \caption{Our TNO catalog of proper semi-major axis, eccentricity, and inclination calculated by SBDynT, zoomed in on the region near the Classical belt. Labels indicate the different regions of the Classical belt, with boundaries defined by the major mean-motion resonances with Neptune, which we show as vertical lines. Note that many of these may not be ``proper'' elements which are formally only defined for non-resonant object; the same calculation that removes forcing frequencies by the planets is often referred to as the ``mean'' elements \citep[see details in ][]{Spencer:2026}. TNOs are colored by the Distance Metric uncertainty which describes the variability of proper elements over the 552 MYr integration \citep[see text and ][]{Spencer:2026}. Cooler colors correspond to non-resonant quasi-periodic orbits while warmer colors indicate orbits affected by mean-motion resonance (some of which are labeled), secular resonance, choas, or scattering (e.g., the highest eccentricities). The use of proper elements highlights an empty and/or highly variable region near proper inclinations of 4-8$^{\circ}$ and between proper semi-major axis of about 42-47 au which is due to a secular resonance as discussed in detail in Section \ref{sec:gs8_res}.}
    \label{fig:small_catalog}
\end{figure*}

We use SBDynT (the Small Bodies Dynamics Tool, V1.0, publicly available at \url{https://github.com/small-body-dynamics/SBDynT}) to calculate proper orbital elements for the entire catalog of known multi-opposition TNOs contained within the JPL Small Body Database as of 04/2026 \citep{Spencer:2026}. 
\citet{Spencer:2026} demonstrated that SBDynT produced proper elements consistent with previous catalogs in accuracy and precision. 
This catalog is partly shown in Figure~\ref{fig:small_catalog}, where we have specifically narrowed the figure to show the Classical belt region. 
The catalog in its entirety is available alongside this text in a machine readable file format.
TNOs are colored in these figures corresponding to the time variability of their proper elements. 
These variations -- called ``uncertainties'' but not related to observational precision -- are then converted into a single Distance Metric using the formula that estimates the velocity difference between objects with similar orbits \cite{Zappala:1990, Spencer:2026}.

\begin{table*}[]
\begin{tabular}{@{}lll@{}}
\toprule
Column name   & Units    & Description                                                                                                             \\ \midrule
Objname*       &         & Primary designation provided by JPL Horizons                                                                  \\
OscSMA        & au      & Osculating Semi-major axis in the Barycentric Coordinate Frame                                                          \\
OscEcc        &         & Osculating Eccentricity in the Barycentric Coordinate Frame                                                             \\
OscInc        & rad & Osculating Inclination in the Invariable Plane                                                                          \\
OscAOP        & rad & Osculating Argument of Periapse $(\omega)$ in the Invariable Plane                                                      \\
OscLAN        & rad & Osculating Longitude of Ascending Node $(\Omega)$ in the Invariable Plane                                               \\
MeanSMA       & au      & Mean Semi-major axis in the Barycentric Coordinate Frame                                                                \\
MeanEcc       &         & Mean Eccentricity in the Barycentric Coordinate Frame                                                                   \\
MeanSinI      &         & Mean Sine of the Inclination in the Invariable Plane                                                                    \\
PropSMA       & au      & Proper Semi-major axis                                                                                                  \\
PropEcc       &         & Proper Eccentricity                                                                                                     \\
PropSinI      &         & Proper Sine of the Inclination                                                                                          \\
PropAOP       & rad & Proper Argument of Periapse $(\omega)$ at epoch t=0                                                                     \\
PropLAN       & rad & Proper Longitude of Ascending Node $(\Omega)$ at epoch t=0                                                              \\
PropSMA\_err  & au      & Numerical Uncertainty of Proper Semi-major axis Computation                                                             \\
PropEcc\_err  &         & Numerical Uncertainty of Proper Eccentricity Computation                                                                \\
PropSinI\_err &         & Numerical Uncertainty of Proper Sine of Inclination Computation                                                         \\
g("/yr)       & "/yr    & The Proper Apsidal Precession rate of the Small Body                                                                    \\
s("/yr)       & "/yr    & The Proper Nodal Precession rate of the Small Body                                                                      \\
g\_err        & "/yr    & Numerical Uncertainty of Proper Apsidal Precession Rate Computation                                                     \\
s\_err        & "/yr    & Numerical Uncertainty of Proper Nodal Precession Rate Computation                                                       \\
data\_arc*     & days    & The length of the observational arc used to compute the orbit                                  \\
H*             &         & Absolute magnitude of the small body  \\
spkid*         &         & The SPK-ID identifier used by the JPL Horizons database                                      \\ \bottomrule
\end{tabular}
\caption{Table of TNO proper elements for the catalog of known objects presented in this paper. The primary dataset used in this paper is the ``AllTNOs\_500myr\_pe'' dataset, which is included in a machine readable format. The table of proper elements computed for the synthetic dataset described in Sections \ref{sec:sec_mapping} has the same format, but naturally do not include the columns marked by an asterisk. The catalog files can be found in the Zenodo repository with this DOI: \url{10.5281/zenodo.22285391}. \textcolor{red}{Edit for Preprint: The Zenodo repository is not active until the review by AAS Journals is complete. In the meantime, these datasets can be found in a GitHub repository associated with this preprint release:} \url{https://github.com/dallinspencer/TNO-Secular-Dynamics-Catalog}}
\label{tab:columns}
\end{table*}

We note that, as mentioned in \cite{Spencer:2026}, there exist long-period secular frequencies which lie on the order of 66 Myr, and produce fractionally large-amplitude terms in the eccentricity and inclination evolution of the low-inclination Cold Classical TNOs. 
The default behavior of SBDynT is to calculate synthetic proper elements from a 150 Myr integration, and to then compute the numerical uncertainty using 5 overlapping windows of 50 Myr each, (75 Myr forwards and backwards), which the uncertainty being represented by the standard deviation between these 5 windows. 
However, this default SBDynT integration for TNOs causes the length of time in each individual window to be shorter than a full circulation period of this secular frequency, which causes the Cold Classical TNOs to have unphysically high uncertainties, while the reported proper elements remain accurate.
To reduce the uncertainties of the Cold Classical TNO, we instead integrate our catalog of TNOs for 500 Myr, rather than the default 150 Myr.

For non-resonant particles, proper elements are essentially conserved quantities (one reason they are also called ``free'' elements), so this distance metric is generally quite small ($\lesssim$ 10 m s$^{-1}$), corresponding to cooler colors in Figure~\ref{fig:small_catalog}. 
Warmer colors thus correspond to particles with more perturbed orbits due to mean-motion resonance, secular resonance, scattering, and/or chaotic evolution. 
See \citet{Spencer:2026} for more details.  

The primary dynamical groups of the Classical belt are very quickly identifiable in Figure~\ref{fig:small_catalog}: (1) A denser population of dynamically ``colder'' TNOs at low inclinations $(I_{free}\lesssim4^{\circ})$, termed the ``Cold Classical TNOs'' (herafter CCTNOs), and (2) a dynamically ``hotter'' and less dense population at higher inclinations termed the Hot Classical TNOs (HCTNOs). 
In addition, there is a visible overdensity of TNOs among the CCTNOs from $43.7<a<44.5$ au at $e_{free}<0.1$, which has been termed the Cold Classical belt ``kernel'' \citep{Petit:2011}.
These dynamical groups are particularly well-defined in the proper elements, which better highlight the edges of the groups in proper semi-major axis, eccentricity, and inclination.

Some particularly sharp features related to the shapes of these populations in the proper elements have also been discussed in earlier studies, including the following.
\begin{enumerate}
    \item The drop-off in the density of the Classical belt TNOs from $42.2<a<43.7$ at $e>0.06$, and a similar density reduction of the kernel at $e>0.1$. 
    This feature was first pointed out and studied in \cite{Dawson:2012}, with \cite{Morbidelli:2014} describing how sweeping mean-motion resonances during Neptune's migration can naturally form these boundaries. 
    While the kernel has been known to be truncated in eccentricity, this cut is quite enhanced in proper element space.
    \item The ``wedge'' shape of increased eccentricity which begins with the kernel, and extends up to the 2:1 mean motion resonance at 47.8 au \citep{Batygin:2011}. 
    The wedge shape has been primarily identified as an underdensity of Classical belt TNOs at low $e$ from 43.7 - 47.8 au, but we point out the underdensity of Classical TNOs at $e>0.1$ in the same region of $a\approx45$ au as well. 
    \item The sharp boundary of the Cold Classical belt at $I=4^{\circ}$, which is bordered by a string of ``unstable'' particles from $4^{\circ}<I_{free}<10^{\circ}$, which we will refer to in this paper as the ``Warm'' Classical belt \citep{VanLaerhoven:2019, Gladman:2021, Huang:2022}.
\end{enumerate}

Many of these features are fossils of solar system evolution, and provide clues to the interactions that occurred as Neptune arrived at its present orbit. 
Others are features of present day instability and/or secular perturbation, which are often highlighted by the larger values of the Distance Metric stability indicator.

To be fully equipped to understand the resulting structures mentioned above, we perform an analysis of the current secular structure of the present outer Solar System in the next section.

\section{Secular Behavior Among the Classical Belt TNOs}
\label{sec:sec_mapping}
 
The long-term orbital evolution of stable, non-scattering, non-resonant objects is primarily driven by their secular precession frequencies.
These frequencies naturally result from the solar system's eigensystem of secular modes produced by interactions amongst all of the planets, primarily the giant planets. 
Thus, the secular precession rates of the solar system small bodies are a function of their location within the solar system (i.e., the present-day osculating orbital elements, excluding mean anomaly).
This dealignment and realignment drives the secular evolution of a TNO's orbit over millions of years (or longer). 

The longer a small body's orbit remains aligned with a perturber, the more strongly the forced terms associated with that planet apply to the orbit; this which is why secular resonant behavior can shape the distribution of small body orbits so strongly. 
Secular resonances occur when the proper precession frequencies $g$ and $s$ are equal to some linear combination of forced planetary frequencies, $g_i$ and $s_i$. Formal secular resonances satisfy two criteria, known as the d'Alembert criteria \citep{Hamilton:1994, SSDBook:1999}.
The first criterion requires the sum of the coefficients for the resonant terms to equal 0 and arises from the fact that resonances are a physical phenomenon that do not depend on the coordinate system. 
The second requires the sum of the nodal precession coefficients, $s,s_i$, to be even which results from the two-fold symmetry of evolution around the total angular momentum plane.
Examples of valid secular resonance combinations would thus include $g-g_8$, $s-2s_7+s_8$, and $g+s-g_8-s_8$; invalid examples include $g-s_8$, $g+s-2g_8$, or $g+s+g_8+s_8$.

Particles near mean-motion resonances are strongly affected on short timescales, with additional forced terms which modify their secular precession frequencies \citep{Malhotra:1989}. 
As a result, secular resonances interact with mean-motion resonances and vice-versa, adding complexity to the secular frequency distribution throughout the solar system. 
Complex and/or overlapping resonant interactions introduce chaotic evolution, which can manifest as both ``resonance sticking'' -- where TNOs tend to spend longer associated with resonances -- and instability -- where TNOs tend to spread quickly through proper element phase space. 

In addition, noticeable discontinuities in the secular frequency distribution can often highlight the impact of even higher-order resonances, both mean motion and secular, which may not necessarily result in the destabilization of particles, but does cause the secular behavior of the TNOs to change.

To map the secular frequency distribution of the outer solar system, we compute the proper orbital elements, including the proper precession frequencies, for a large set of synthetic TNO Classical belt particles. %, and show the results in this section. 
This allows us to identify regions of secular resonance or near resonance and make comparisons to maps of the stability for these same particles, providing new insights into how these resonances contribute to stability or instability in these regions.

\subsection{A Secular Mapping of the Classical Belt TNOs}\label{s:secular_map}

Characterizing the secular behavior of TNOs across a given region of phase space can broadly be approached in one of two ways, (1) analytically, through expansion of the disturbing function to some chosen order \citep[such as has been done by][]{Milani:1994, Knezevic:2000, Knezevic:2003}, or (2) numerically, through direct integration of a grid of test particles, and studying their mean behavior.
Analytical approaches offer elegant closed-form descriptions of secular dynamics, but their accuracy is inherently limited by the truncation order.
Analytical methods can thus often struggle when considering regions of orbital phase space which experience strong higher-order terms, such as near mean motion resonances or at large eccentricity and/or inclinations \citep{Knezevic:2019}. 
Numerical integration and computation of the proper elements, such as by way of SBDynT which filters out perturbations in frequency space \citep[see full details in][]{Spencer:2026}, can naturally capture some higher-order effects without requiring explicit expansion.
This method also carries the advantage of encoding dynamical stability directly into the resulting secular map; particles that are removed by instability simply do not contribute to the mean behavior of the region.
For these reasons, we adopt the numerical approach here.

We first initialize a \textsc{rebound} \citep{Rebound:2012} simulation with the Sun and the 4 giant planets using SBDynT (the masses of the terrestrial planets are folded into the Sun) at the epoch 2456220.5 TDB, which is the epoch corresponding to the orbital solution returned by JPL Horizons for the TNO (15760) Albion.

We then implant a grid of 400,000 Classical Belt TNO particles uniformly sampling the following initial osculating barycentric orbital parameter ranges: $40 < a < 48$ au, $0 < e < 0.3$, and $0^{\circ} < I < 40^{\circ}$, with the $\varpi$, $\Omega$ and $M$ angles being defined randomly from 0 to $360^\circ$.
The inclinations and orbital angles of the particles are defined with respect to the invariable (e.g., total angular momentum) plane of the initialized planetary system.

We then integrate this grid of particles using the \textsc{mercurius} integrator \citep{Reboundmercurius:2019} for 4.5 Gyr, and then compute the proper orbital elements using the evolving osculating elements of the small body during the final 3.5 Gyr of the integration for particles that survive the entire simulation in the classical belt. 
The first Gyr of the integration serves effectively as a form of burn-in, which removes most of the unstable particles from the simulation, since proper elements are not well-defined for unstable particles.
The final grid of TNO proper elements can be used to find the mean behavior of TNOs with similar proper elements. 
We demonstrate how this grid of particles may be used to better understand aspects of the Classical belt TNOs.

\subsection{Stability and Long-term Occupation of Classical TNOs}
\label{sec:stability_maps}

\begin{figure*}[ht]
    \centering
    \includegraphics[width=1\linewidth]{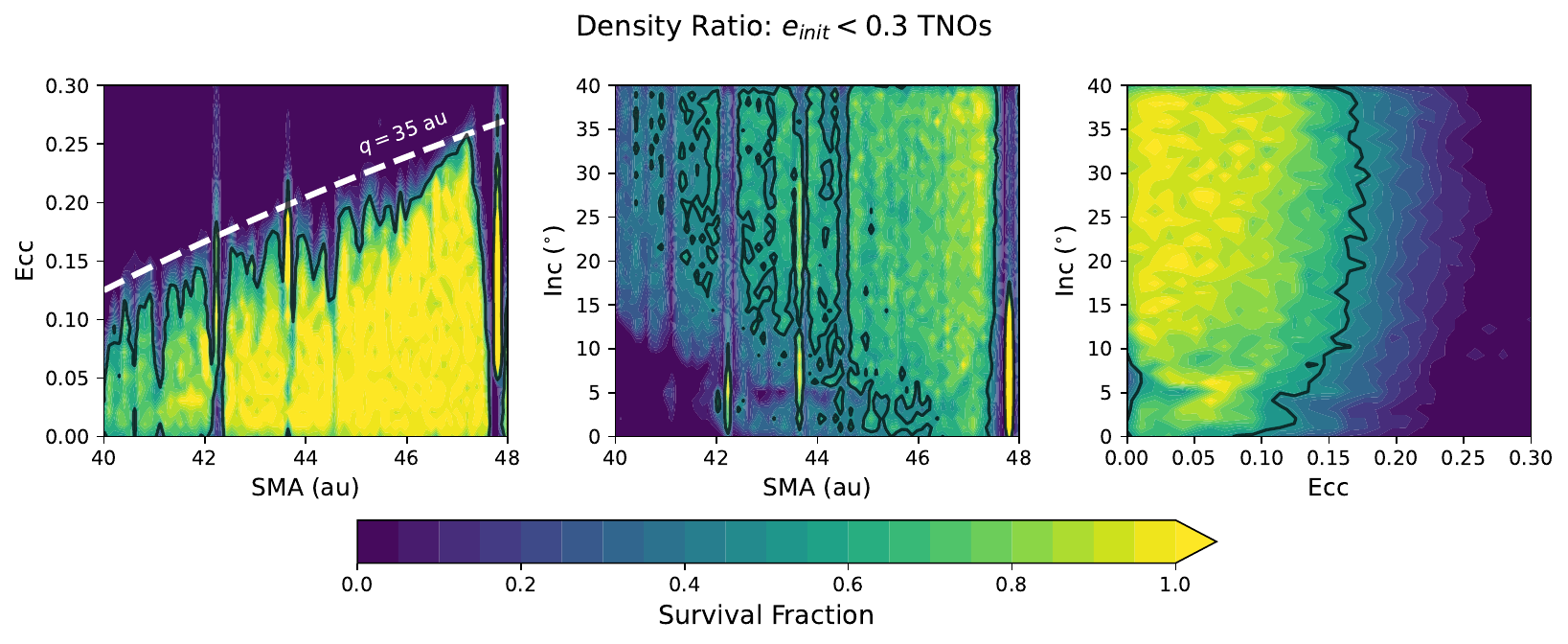}
    \includegraphics[width=1\linewidth]{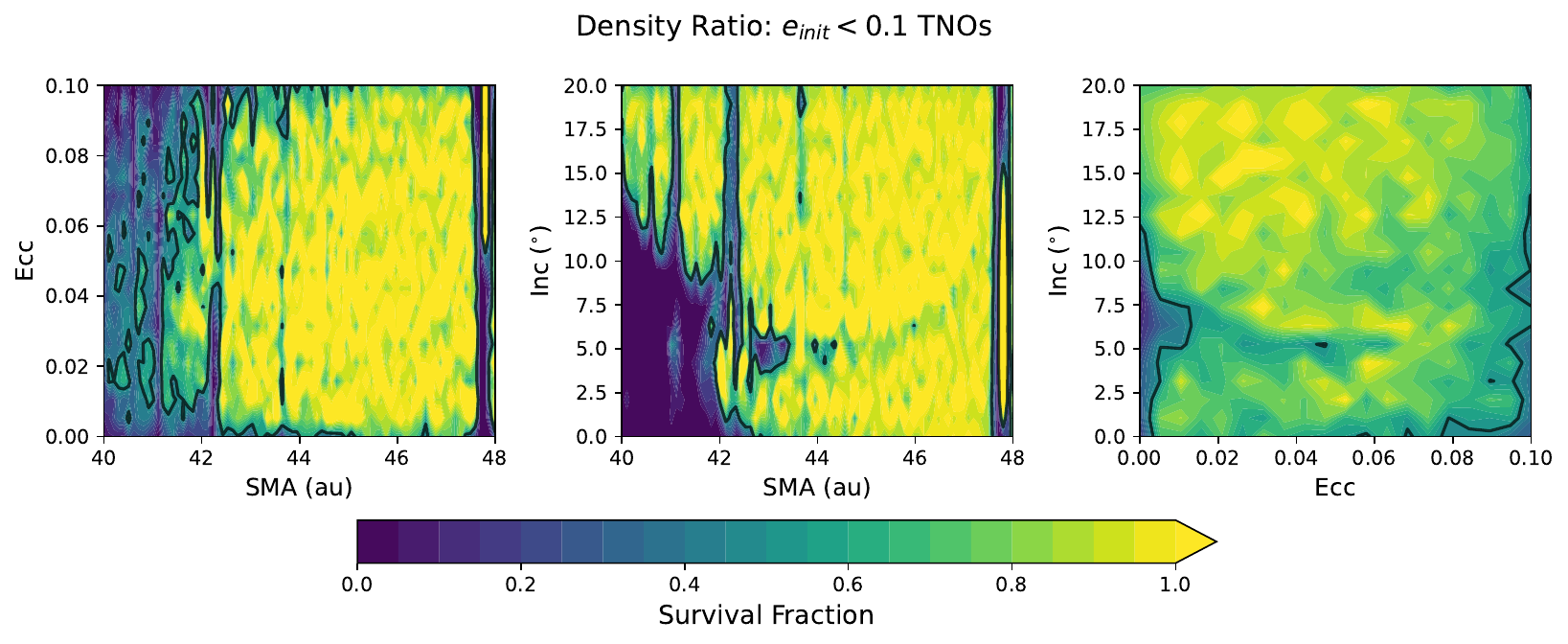}
    \caption{Top Panel: The density of particles in proper $a$, $e$, $I$ parameter space after 4.5 Gyr of integration. The colorbar represents the fraction of particles in the contour, normalized as a survival fraction of the initial population of objects in that bin. The overdensities produced by mean-motion resonances at islands of stability are clearly visible in the proper element space. In addition, the instabilities associated with large eccentricities or secular resonances are also visible. Regions of the contour maps extend above 100\% as some regions, such as mean-motion resonances, pull nearby particles to stable islands at the resonant center in proper element space; regions with a survival fraction greater than 100\% are simply clipped to the maximum value on the colorbar.
    A dark contour line indicates where the resulting survival fraction of particles transitions to less than 50\%.
    Bottom Panel: The density of particles with $e_{init}<0.1$ in proper a,e,I parameter space after 4.5 Gyr of integration. The cut at $e<0.1$ produces a stability map where instabilities are now primarily a function of semi-major axis and inclination, revealing inclination instabilities within the Classical belt.}
    \label{fig:density_hist}
\end{figure*}

Perhaps the most common use of computing a grid of synthetic particles is to measure the stability of the small bodies in that particular region of orbital parameter space.
Examples of this method can be found in many previous works studying different small body populations, such as for Centaurs, Classical belt TNOs, and resonant TNO populations \cite[i.e.]{Kuchner:2002, Tiscareno:2003, Lykawka:2005, Tiscareno:2009}.
While stability can be measured and defined in a multitude of ways, generally the first test of stability for a region in phase space is to measure the survival fraction of simulated particles which have been implanted in a certain region of phase space, which remain within a reasonable distance of their initial osculating elements by the end of the simulation, and haven't been scattered into the inner Solar System ($a_{fin}<30.1$au), or ejected entirely ($e_{fin}>1$).
This is typically done in the osculating element phase space, but this type of analysis can be heavily improved by considering the density of stable orbits in proper element space.
We show this resulting stability map in Figure~\ref{fig:density_hist} for two different eccentricity cut-offs; both maps show very clear features of stability and instability in the Classical belt. 

Figure~\ref{fig:density_hist} clearly shows the strong correlation between $q$ and the long-term stability of particles; objects with $q < 35$ au are almost always scattered and/or removed by Neptune, which is well-known and documented \citep[e.g.][]{Duncan:1995}.
In addition, the impact of mean-motion resonances becomes instantly noticeable, with vertical lines showing regions of unstable parameter space and/or stable parameter space within the mean motion resonances. 

While informative, the shape of visible inclination-dependent instabilities are less defined in the top-center a vs I panel of Figure~\ref{fig:density_hist}, due to the large number of unstable particles found at high-e. 
Indeed, more than half of all of the Classical belt particles at the inner portion of our grid are unstable, since the eccentricity boundary occurs near $e\approx0.15$, and most of these TNOs are lost to close encounters on relatively short timescales. 
When all particles are included in the normalized density map, this eccentricity-instability projects into the $a$-$I$ plane, which is not necessarily related directly to the inclination-dependent dynamical structure.

In addition, the lower range of eccentricity is more representative of the Cold Classical belt TNO population, which is of interest to the broad planetary science community.
To better display the inclination-space instabilities, the bottom panel of Figure~\ref{fig:density_hist} restricts the sample to particles with initial eccentricities $e_{init}<0.1$, which is largely stable across the semi-major axis and eccentricity span of interest. 
It becomes immediately clearer in the bottom panel that there exist significant inclination instabilities at $a< 45$ au, specifically (1) a strong feature at $a_{free}<42$ au, and $I_{free} \lesssim 8^{\circ} - 15^{\circ}$, and (2) a smaller feature which occurs at $42.3<a<44.5$ au and $4^{\circ}<I_{free}<6.5^{\circ}$.
This first feature is particularly well-documented, and is caused by the $\nu_{18}$ linear secular resonance which occurs when $s = s_8$, which is equivalent to $\Dot{\Omega} \approx \Dot{\Omega}_{Nep}$  \citep[see, e.g., early works by][]{Knezevic:1991,Holman:1993, Duncan:1995}.
The $\nu_{18}$ resonance may have played an important role in depopulating the massive planetesimal disk, contributing to Neptune's outwards migration, and producing our present day solar system orbital configuration \citep{Levison:1998}.

The second feature is less documented in the literature.
The earliest discussion of the second notable secular feature at moderate proper inclinations we find is by \cite{Kuchner:2002}, who integrated the orbits of 1458 implanted Classical belt particles and measured stability in different regions of phase space. 
At the time, they recognized that the web of secular resonances was complex, and determined that the instability was caused in most part by overlapping $\nu_{17}$ and $\nu_{18}$ secular resonances, because this was primarily a feature seen only in inclination space. 
Later, \cite{Lykawka:2005} produced stability maps of the Classical belt TNOs and came to the conclusion that instead this instability was primarily related to many overlapping mean motion resonances between Uranus and Neptune, as it is well known that many higher order resonances do exist at these semi-major axes \cite[see, e.g., a recent compilation by][]{Smirnov:2025}. 

More recently, \cite{Huang:2022} identified this region of phase space as displaying both a sharp underdensity in the number of observed TNOs as well as exhibiting generally larger forced secular terms, leading to poor estimations of the proper elements for these TNOs. 
Consistent with \cite{Huang:2022}'s result, we find that this feature is notably visible in Figure~\ref{fig:small_catalog}, with nearly every TNO in the range $4.5^{\circ} < I_{free}<8.5^{\circ}$ displaying large distance metric indicators, reflecting increased numerical uncertainty of the proper element calculation due to significant forced terms on these particle orbits.

The shortcomings of all of these previous analyses is that (1) the sample sizes of investigated particles in the region of interest were too small to properly find the shape of this instability, and (2) previous works have not computed the proper orbital elements for their simulated population, where this instability is most clearly defined.
It has been noted that this instability was generally limited to $4^{\circ}<I_{free}<10^{\circ}$, but the clear shape of this instability has not been seen until now with our much denser grid of TNO particles with their computed proper elements.

While measuring the density of particles gives some insight into the instability of the region, it is also helpful to measure how the orbits of remaining particles are affected, to gain insight into what may be causing this instability.
We do this using stability indicators, which can help identify regions of chaotic or large-amplitude evolution, such as is provided by SBDynT \citep{Spencer:2026}. 

\begin{figure*}[ht]
    \centering
    \includegraphics[width=1\linewidth]{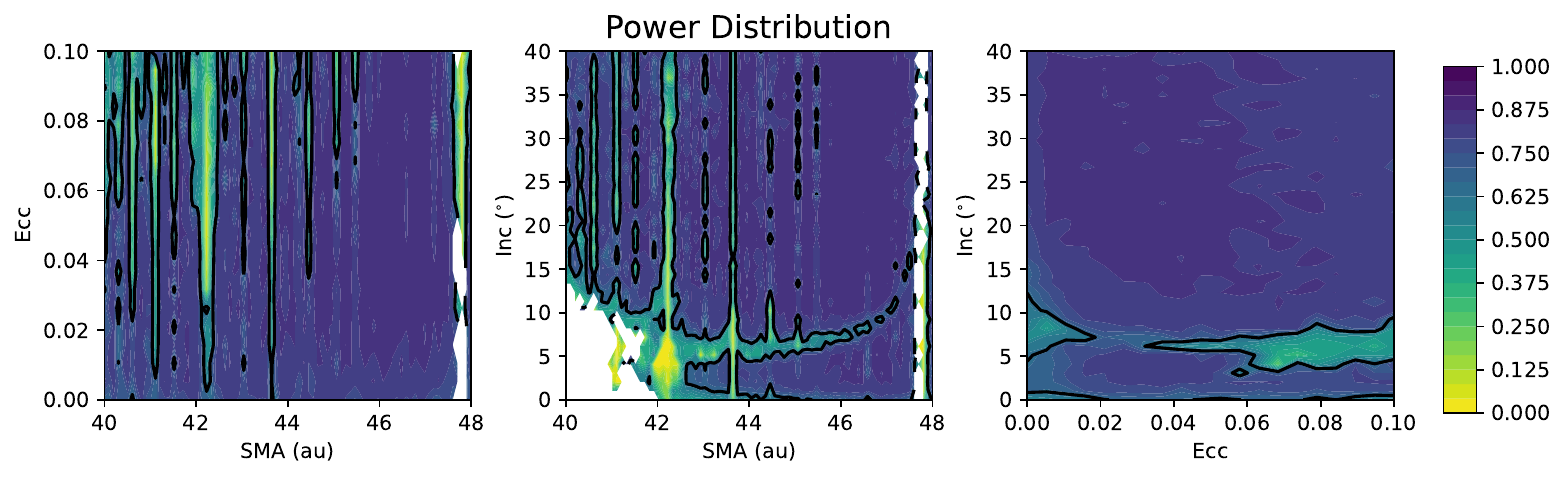}
    \includegraphics[width=1\linewidth]{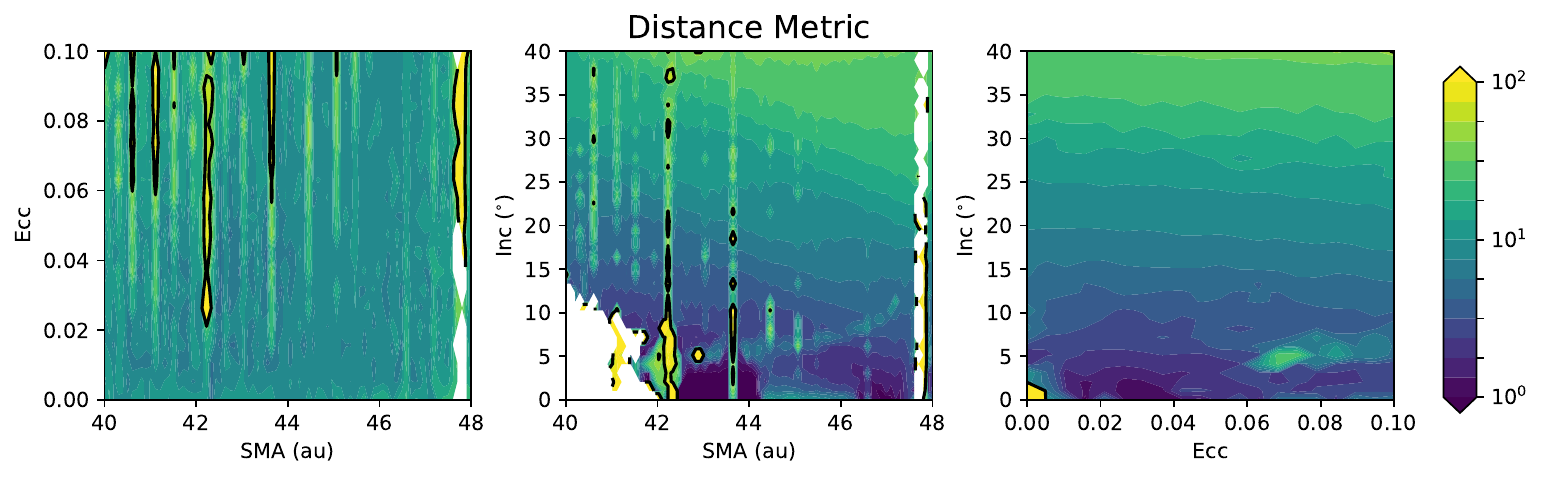}
    \caption{Stability maps of the Power Distribution and Distance Metric stability indicators provided by SBDynT. ACFI and Entropy indicators are not shown, as they are not well-behaved at the higher resolution used in our analysis. Contours are shown indicating the boundary of SBDynT's stability criterion for each indicator (Distance Metric = 50 m/s, Power Distribution = 60\%). A clear feature is seen in $a,I$ space for both stability maps between $5^{\circ} < I_{free} < 8.5^{\circ}$, indicating instability which is caused by large forced terms in TNO orbital evolution in this region of proper element space. }
    \label{fig:stability_maps}
\end{figure*}

We compute stability indicator maps for the distance metric indicator and the power distribution indicator provided by SBDynT, shown in Figure \ref{fig:stability_maps}, and restrict the eccentricity range like we did in Figure \ref{fig:density_hist} to limit the impact of high eccentricity on displaying the inclination-dependent instabilities.
While SBDynT provides access to additional stability indicators, like the Autocorrelation Function Index (ACFI) or the Information Entropy indicator of the orbit's angular momentum, these other indicators are very sensitive to the time resolution of the time arrays, while the distance metric and power distribution indicators are actually improved by longer integration despite lowered time resolution. 

The moderate-inclination feature near the middle of the classical belt is quite noticeable in both stability maps in Figure-\ref{fig:stability_maps}.
Interestingly, the feature actually appears to extend to higher semi-major axes as well, which is not seen in Figure-\ref{fig:density_hist}, since the instability is not strong enough to actually remove particles from this region. 
The consistency of the secular forcing across such a wide and continuous range of semimajor axes implies that this instability does not result directly from mean-motion resonance, but is likely related to a secularly resonant effect, as was suggested by \cite{Kuchner:2002}.
Further investigation of this requires an analysis of the secular frequency distribution throughout phase space.

\subsection{Secular Frequency Maps}
\label{sec:sec_freq_maps}

Secular resonances, while extremely relevant to the formation and sculpting of small body populations, are often difficult to represent or describe due to their complexity in orbital parameter space.
Indeed, at even the simplest approximation, secular resonances are still 3-dimensional functions of an object's semi-major axis, eccentricity, and inclination; further complexity is added by the $\omega$, $\Omega$ angles of the object's orbit.
Thus, calculating and visualizing the structure of secular resonances can be very complex.

Previous studies have used both analytical and numerical methods to visualize secular resonances in the asteroid belt and the transneptunian region \cite{Knezevic:1991,Milani:1994, Knezevic:2003}. 
However, analytical methods are often limited to 4th-order approximations due to the complexity of additional orders of magnitude.
Numerical methods are in some ways even more constrained, as successfully representing the smooth function of the full 5-dimensional parameter space of the $g$ and $s$ frequencies requires the integration of a sufficiently large sample of particles to adequately measure the precession frequencies across the full parameter space; this can be computationally expensive, especially for large ranges in semimajor axis. 
However, understanding the complex web of secular resonances and their interactions with nearby mean-motion resonances is key to understanding the complex dynamics that sculpt the small-body populations of the solar system and, as mentioned earlier, is best achieved with numerical investigations \citep{Knezevic:2019}.

\begin{figure*}
    \centering
    \includegraphics[width=1\linewidth]{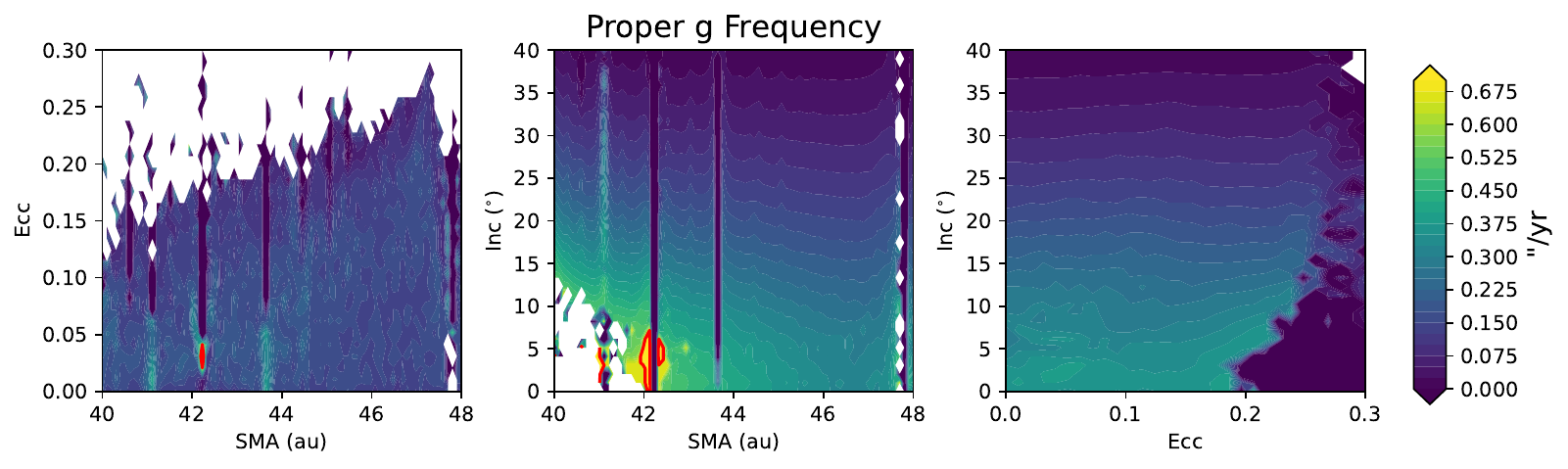}
    \includegraphics[width=1\linewidth]{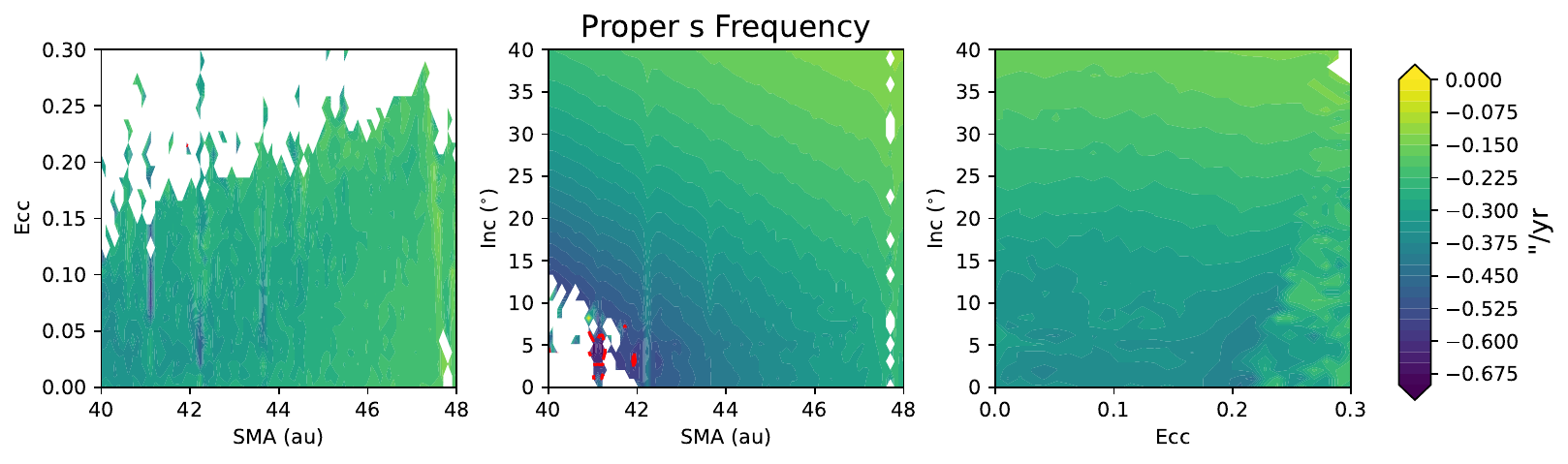}
    \includegraphics[width=1\linewidth]{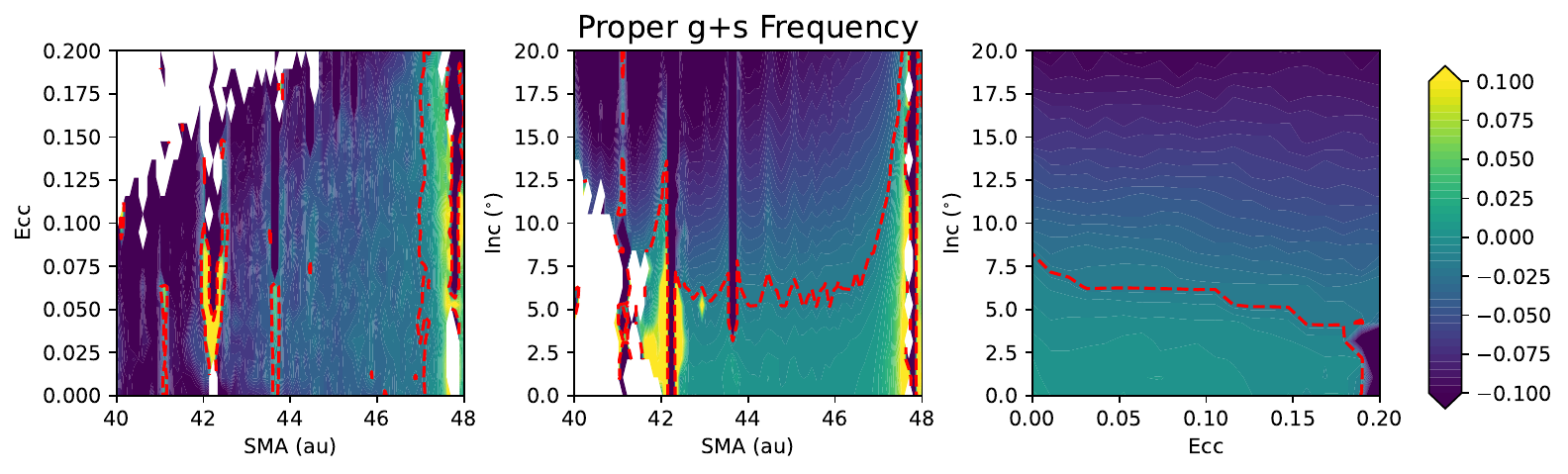}
    \caption{The distribution of proper $g$ and $s$ frequency's throughout proper element space among the Classical belt TNOs with $e<0.3$ and $I<40^{\circ}$. Red contours indicate the location of the Neptunian secular resonance associated with that secular frequency. For example, the red line in the 3rd row indicates the line where $g+s=g_8+s_8$. Note the limits placed on the axes for the $g+s$ figure, to better demonstrate the shape of the resonance at the relevant orbital parameters.}
    \label{fig:prop_freqs}
\end{figure*}

To properly measure this secular frequency distribution throughout the Classical belt, we compute proper $g$ and $s$ frequencies across our large sample of Classical TNO particles. 
The resulting proper frequencies for the full phase space, while 5-dimensional, can be reasonably well represented by producing a contour map of the median proper frequencies for each bin of $a$, $e$, and $I$.
Figure~\ref{fig:prop_freqs} shows the 2-dimensional visualization of the $g=\Dot{\varpi}$, $s=\Dot{\Omega}$, and $g+s=\Dot{\varpi}+\Dot{\varpi}$ precession frequencies across our grid of TNO particles in proper $a$, $e$, and  $I$. 
A red contour is plotted in each figure to indicate the estimated center of the relevant linear secular resonance, i.e. where $g = g_8$ (the $\nu_8$ resonance), $s=s_8$ (the $\nu_{18}$ resonance), and $g+s = g_8+s_8$.

We note that while there is no visible contour for the $\nu_{18}$ resonance at the $a<42$, $I<15^{\circ}$ instability, that is because no particles remain to fill that portion of phase space; it should be clear from the proper $s$ frequency distribution in $a$-$I$ that this secular resonance occurs within this region, as the median secular frequency of surviving particles approaches $s=s_8\approx-0.692 \;\, "/yr$.
The remaining $\nu_8$ and $\nu_{18}$ resonant particles are relegated almost entirely to having orbits near the 5:3 mean motion resonance.
The proximity of this mean motion resonance to the secular resonances at the $e=0, I=0^{\circ}$ bases (see \cite{Knezevic:2003} for analytical examples of the $\nu_8$ and $\nu_{18}$ secular resonance shapes), leaves this as the only space where stable TNOs can have proper motion which is not dominated by Neptune's secular precession rate; the secular forced terms are broken up by the faster mean motion resonance.

In looking at Figure-\ref{fig:prop_freqs}, it becomes immediately apparent that the center of the $\nu_8+\nu_{18}$ resonance lies precisely along the observed inclination instability discussed in Section-\ref{sec:stability_maps}, indicating that the $\nu_8+\nu_{18}$ is likely related to this inclination-dependent instability.

The $\nu_8+\nu_{18}$ resonance has been considered very little in previous works.
\cite{Morbidelli:1995} analytically investigated the Classical belt when only 5 known TNOs had been discovered to date. 
They identified the $\nu_8+\nu_{18}$ resonance as representing a significant instability at $a<42$ au, and contributed heavily to the destabilization of the low-$e$ and low-$I$ particles within the $40<a<42$ au, $I < 15^{\circ}$ inclination instability. 
However, \cite{Morbidelli:1995} stated that this resonance should be inconsequential at $a>42$ au, because the non-linear resonant terms would become far too weak.
\cite{Knezevic:2003} noted that despite the $\nu_8+\nu_{18}$ resonant terms becoming weaker at larger semi-major axes, small $g+s-g_8-s_8$ divisors caused large-amplitude and very long-period oscillation in the eccentricity and inclination evolution for all of the Cold Classical belt TNOs; this result was also discussed recently in \cite{Spencer:2026} as contributing to the larger instabilities measured for the Cold Classical belt TNOs.
This secular frequency is in fact the long-period forced term discussion Section-\ref{sec:prop_catalog} which prompted the longer integration time of 500 Myr for our TNO catalog of proper elements. 

The very close association of the resonant center with the center of the visible inclination dependent instability very strongly indicates that the this resonance does contribute more significantly to the secular evolution of the Classical belt TNOs than has been previously considered.
In the next section, we present and discuss the mechanism by which the $\nu_8+\nu_{18}$ resonance works to destabilize these particles, and the resulting orbits of the particles which remain near the resonance.

\subsection{The $\nu_8 + \nu_{18}$ Resonance and Orbital Evolution}
\label{sec:gs8_res}

Linear secular resonances occur when the secular precession rates of the orbital angles of a small body coincide directly with the giant planets.
For example, the $\nu_8$ resonance occurs when the proper apsidal precession frequency of a small body, $g\approx1/\Dot{\varpi}$ becomes nearly equal to the proper apsidal precession frequency of Neptune $g_8\approx\Dot{\varpi}_8$,  or $g+g_8\approx0$.
Similarly, the $\nu_{18}$ resonance occurs when the proper nodal precession frequency of a small body $s\approx1/\Dot{\Omega}$, is approximately equal to the nodal precession frequency of Neptune, causing $s+s_8\approx0$. 
Because of this, a secularly resonant object can be identified by the evolution of the resonant angles, i.e. $\varpi - \varpi_{Nep}$ and $\Omega - \Omega_{Nep}$.
If the resonant angle does not circulate from $0^{\circ}-360^{\circ}$, but instead librates, this generally indicates that the small body is resonant, since the time evolution of the resonant angle is technically 0.

Secular resonances are generally associated with orbital instability because of how they impact the growth of eccentricity and inclination as defined in the disturbing function.
In the expansion of the disturbing function, the contribution of individual secular frequencies to the evolution of $e$ and $I$ are defined as periodic contributions with amplitudes which are set by the resonant frequency separation $1/\nu_i$, where the $\nu_i$ divisor indicates the specific frequency combination relative to that term.
It is clear then that a resonant frequency separation close to 0 will cause the forced terms in $e$ and $I$ to become very large, or even approach infinity, which generally leads to orbital instability when $e$ values are pumped, causing the orbits to evolve onto planet crossing orbits. 

A non-linear resonance is similar, but consists of higher-order linear combinations of the small body proper frequencies and the planetary proper frequencies, which must still satisfy the d'Alembert criteria discussed previously. 
The $\nu_8+\nu_{18}$ is an example then of a non-linear secular resonance, which occurs when the sum of the apsidal and nodal proper secular frequencies of a small body are equal to the sum of the same Neptunian frequencies, or $g+s=g_8+s_8 \approx -0.017\;"/yr$.
Since this resonance does not correspond only to the nodal and apsidal precession, we define the angle $\phi=\varpi+\Omega$ to represent the precession described by the $\nu_8+\nu_{18}$ resonance, which should precess at rate equal to that of $\Dot{\varpi}+\Dot{\Omega}$.

We point out that $g_8+s_8$ frequency is the longest period secular frequency among the solar system eigenmodes, with a circulation period of $\approx 66$ Myr.
This means that the resonant term caused by $\nu_8+\nu_{18}$ is generally on the order of hundreds of millions of years for resonant TNOs, and as a 4th order nonlinear resonance, the relevant terms in the evolution of $e$ and $I$ are expected to be very small in most cases.
For these reasons, the $\nu_8+\nu_{18}$ resonance has been considered to be likely irrelevant in the evolution of the Classical belt TNOs, besides those for which $g\approx g_8$ and $s\approx s_8$, such as was discussed by \cite{Morbidelli:1995}. 
However, the very slow $g_8+s_8$ term does not hinder the resonant frequency, but rather enhances it, as the small divisor $(g + s - g_8 - s_8)$ can approach 0 for a wide range of $g$ and $s$ proper frequencies.

\begin{figure*}
    \centering
    \includegraphics[width=1\linewidth]{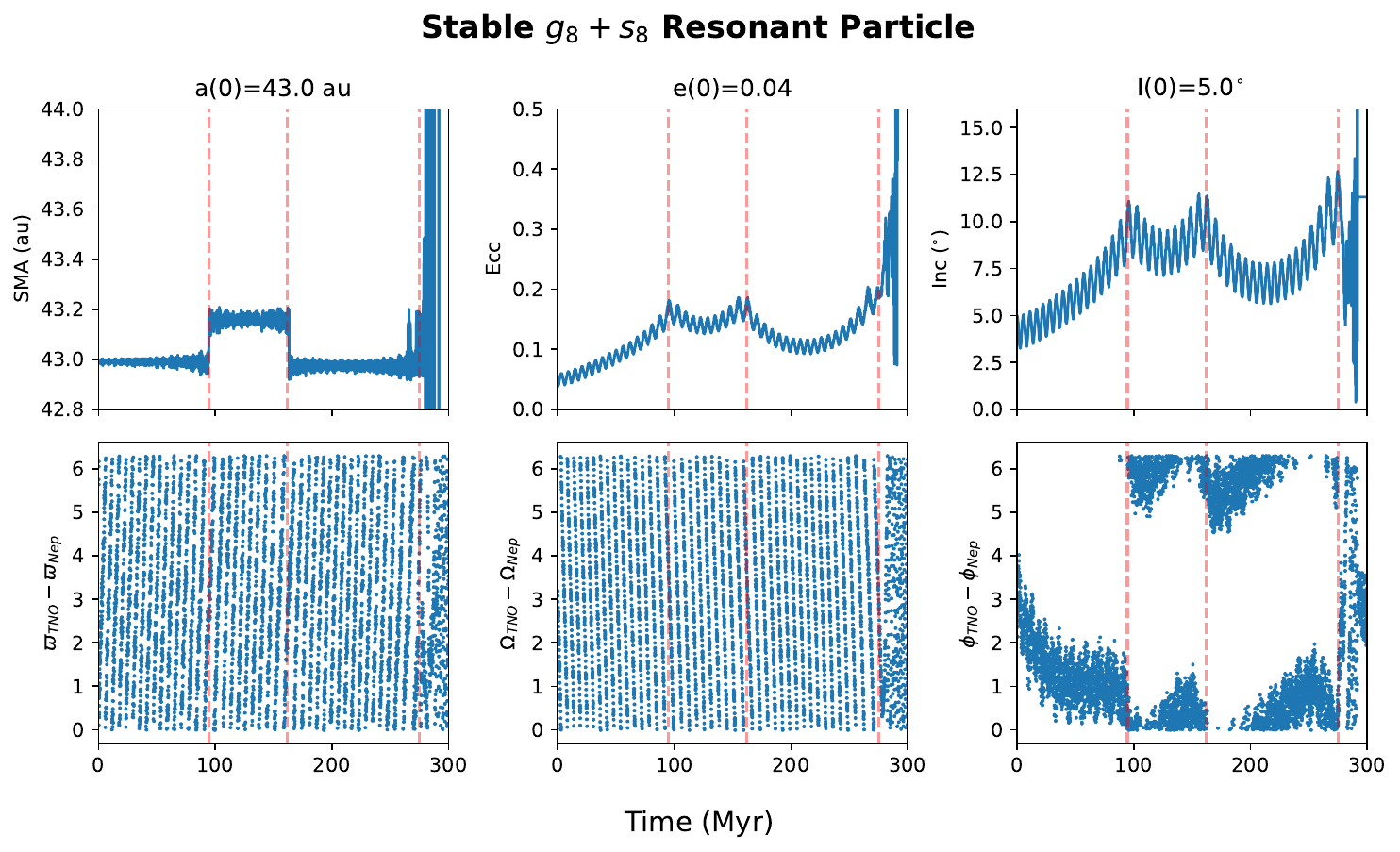}
    \includegraphics[width=1\linewidth]{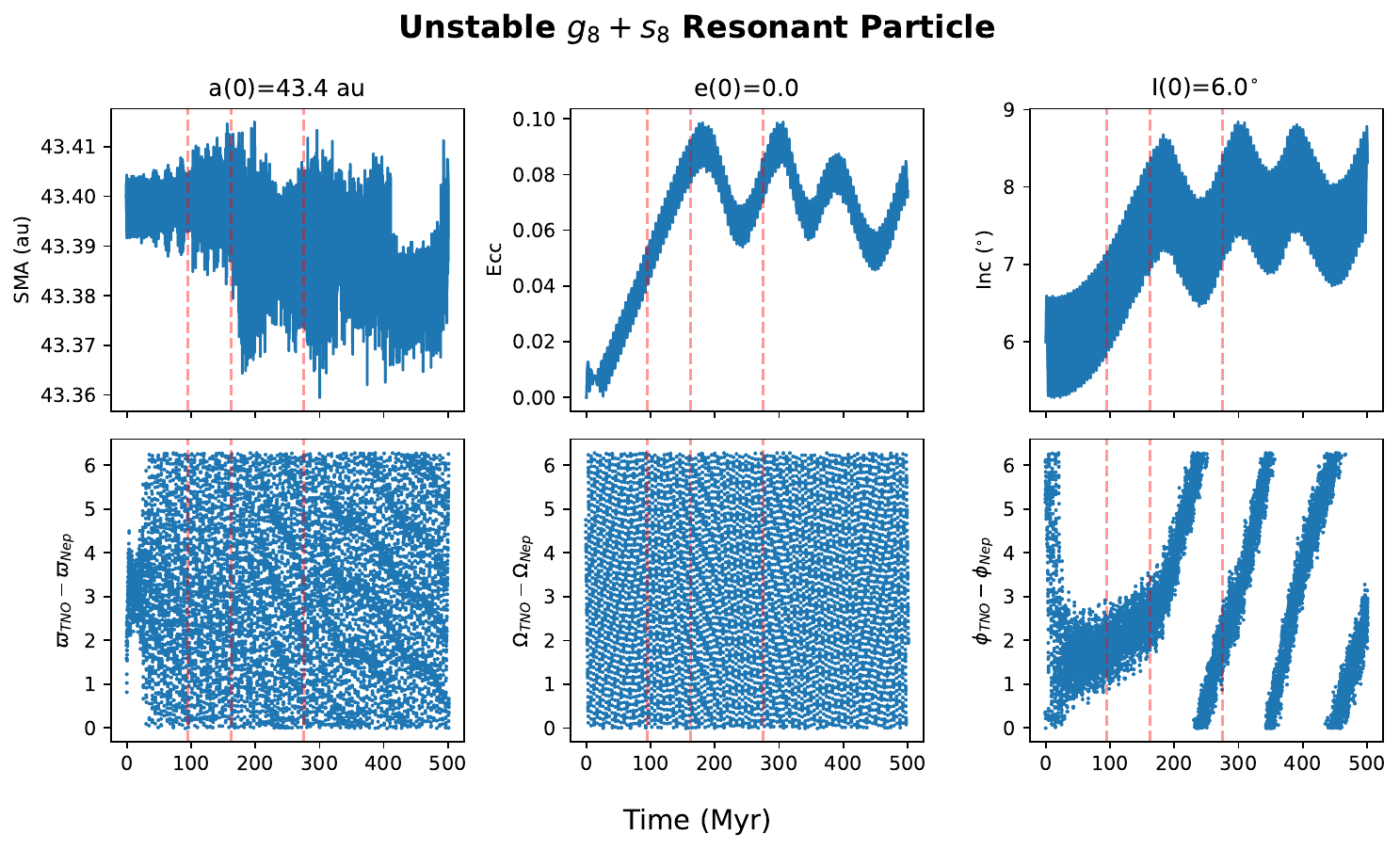}
    \caption{Top: Orbital evolution of a $g_8+s_8$ resonant particle with initial osculating $e=0,\, I=6^{\circ}$, and $g+s\approx-0.021\;"/yr$. Notice that the small initial eccentricity causes the object to originally librate in the $g-g_8$ resonance, which than transitions into libration in the $g_8+s_8$ resonance.  A vertical line indicates where the resonance is broken, allowing the particles to remain stable in new parameter space. 
    Bottom: An unstable $g_8+s_8$ resonant particle with initial $e=0.04$ and $g+s\approx-0.017\;"/yr$. Vertical lines indicate points where $\phi_{TNO} - \phi_{Nep}=0$, causing enhanced forced terms, and causing significant transfer of energy into the TNO orbit by Neptune. }
    \label{fig:sim_gs8}
\end{figure*}

To demonstrate this mechanism, we show two examples of simulated particles in Figure~\ref{fig:sim_gs8} that experience libration in the $g_8+s_8$ resonance. 
In the first case, the particle experiences adiabatic growth in both eccentricity and inclination, until eventually the secular resonance breaks; this initial breaking of the resonance corresponds to a close approach with Neptune due to the enhanced eccentricity of the particle's orbit.
As the eccentricity remains high, mean-motion forcing primarily modifies the apsidal precession rate of the particle, causing the secular resonance to remain broken.
In other words, shorter-period forced terms begin to dominate the orbit, though it is obvious that the $g+s-g_8-s_8$ resonant term continues to be the largest amplitude term in the particle's eccentricity evolution.

In the second case (bottom panels of Figure~\ref{fig:sim_gs8}), we see that the TNO particle similarly experiences adiabatic growth in both the eccentricity and inclination, until the particle is scattered after 100 Myr.
A brief look at the $\Delta\phi$ resonant angle finds that there seems to be a propensity for scattering to occur when $\Delta\phi\approx0^{\circ}$.
The reason for this becomes clear when considering the actual geometry of the resonant angle $\Delta\phi=\Delta\Omega + \Delta\varpi=0^{\circ}$.
A $\Delta \phi=0^{\circ}$ angle occurs when the distance between the orbital apsides is equal to the distance between the orbital nodes, which produces symmetry in the co-occurring orbital configuration.
This symmetry by itself is not generally very significant; however, the particular configurations between orbits which minimize the distance between them require a $\Delta\phi=0$, namely, (1) $\Delta\Omega = \Delta\varpi=0^{\circ}$ and (2) $\Delta\Omega=\Delta\varpi=180^{\circ}$. 
In the first case, the orientation of each orbit is totally aligned, with the line of apsides and the line of nodes for each orbit lying in the same directions. 
In the second case, the line of nodes and the line of apsides are anti-aligned with each other.
This configuration causes the alignment of each orbit's perihelion with the aphelion of the other, and vice versa.

This second configuration is particularly impactful, as it not only produces the smallest possible orbital intersection distance between Neptune and the particle, increasing the probability of and impact of scattering with Neptune, but it also maximizes the difference in the z-velocity between Neptune and the particle, causing the induced torque by Neptune on the TNO orbit to be be maximized.

This first scattering event transfers significant energy into the TNO's orbit, causing a $\Delta a = +0.2$ au, completely changing the secular resonant structure.
However, the $\nu_8+\nu_{18}$ resonant term remains significant, and after a full 66 Myr circulation by Neptune, the particle is scattered again, seemingly back to the original resonant structure, until it experiences a final scattering $\approx132$ Myr later, when the particle is finally ejected from its orbit.

It is worth considering the geometric implication of the $\nu_8+\nu_{18}$ resonance and the accompanying angle $\phi$ with respect to the particle orbits.
It is simple enough to picture and describe the geometric interpretation of the linear secular resonances, $g-g_8$ and $s-s_8$, which indicate that the apsidal precession and nodal precession rates are equal with Neptune. 
However, $\nu_8+\nu_{18}$ would indicate the sum of these precession rates being equal, which does not have a direct analog in the secular representation of an orbit.
A geometric analog which corresponds to the $\phi$ angle can be found by selecting some singular point on the orbital ellipse, and tracing out the evolution of this point over time.
A natural selection for this point would be either the point of perihelion and aphelion on the orbital ellipse, whose positions are defined specifically with respect to the orbital shape and orientation. 

We can thus consider the trace of the periapse or apoapse in 3-dimensions over time as the jointly measured precession of the TNO apsides and nodes over time.
In the case of an object with a $g+s=0\; "/yr$ term, the shape traced out by the periapse represents a hyperbolic paraboloid.
If the $g+s$ is simply close to zero, then the periapse traces out the same hyperbolic paraboloid shape, which now rotates itself around the vertical z-axis of the invariable plane.
The rotation period of the hyperbolic paraboloid around the z-axis is then equal to $2\times 1/(g+s)$, or two times the circulation period for that frequency term.
We note that due to symmetry, this results in an alignment of the two shapes every half a rotation, equivalent to the circulation period itself.

\begin{figure}[!ht]
    \centering

    %\hspace{-1.1cm}
    \includegraphics[width=1\linewidth]{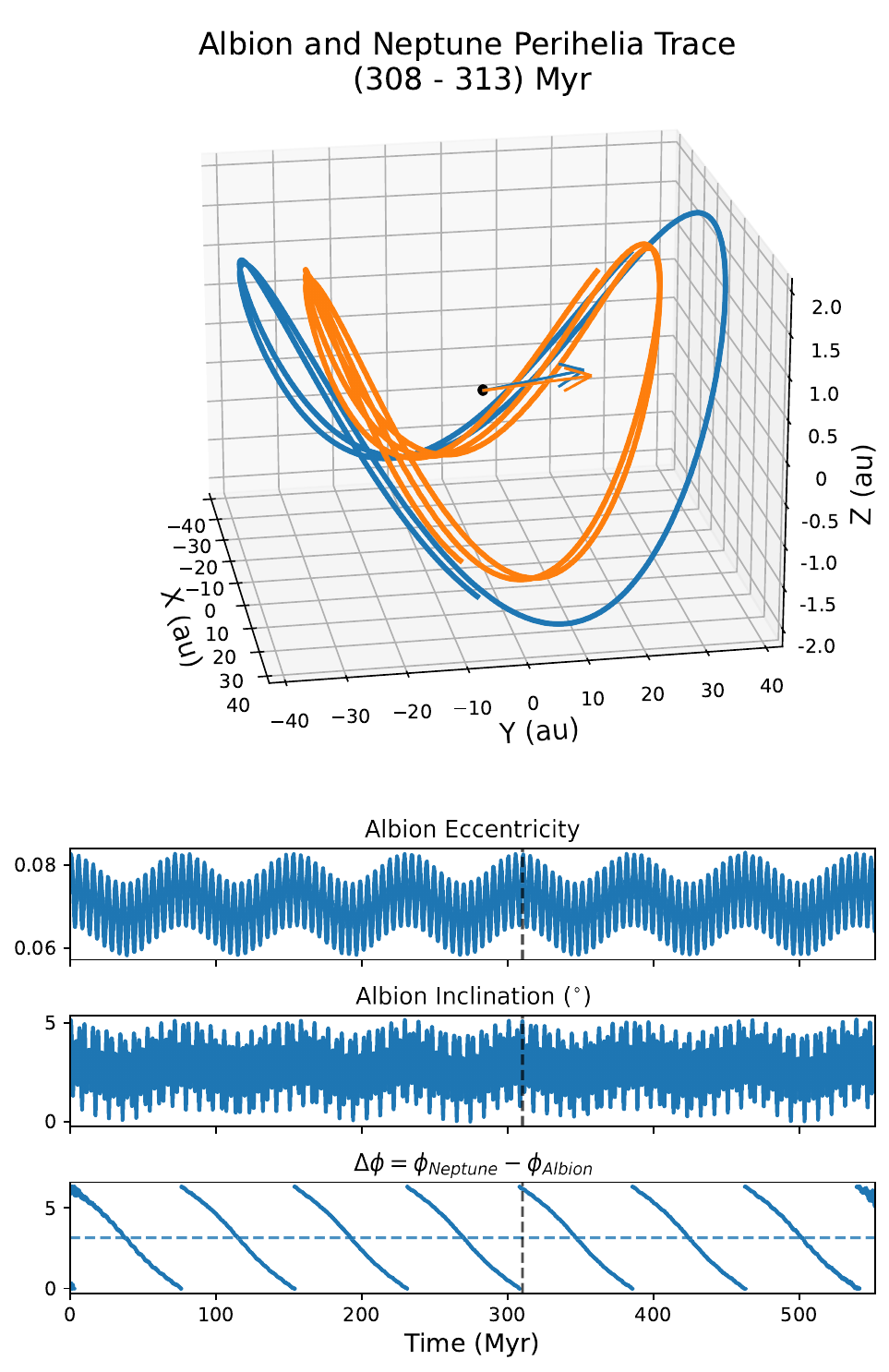}
    \caption{The trace of periapse (q) for TNO (15760) Albion (blue) and Neptune (orange) from t=[308, 313] Myr in the barycentric coordinate frame. An arrow demonstrates the mean $\phi$ angle for each orbit in the frame. The alignment of the traced hyperbolic paraboloid shapes correspond to the alignment between the $\phi$ angles at $\approx 310$ Myr. The z-axis for Neptune is enhanced by 4.5x to better display the nested shape of this orbital configuration.}
    \label{fig:trace_15760}
\end{figure}

An example is shown in Figure~\ref{fig:trace_15760}, which demonstrates the traced periapse in blue for the TNO (15760) Albion and the apoapse of Neptune in orange for a 15 Myr range of a REBOUND integration where the $\Delta \phi$ between Neptune and Albion is close to $0^{\circ}$.
This Figure corresponds to an animation of the evolving interaction between Albion and Neptune over 500 Myr, which we include among our supplementary materials attached to this paper.
This animation makes it clear how the interaction between the rotating hyperbolic paraboloids of the two bodies cause the eccentricity and inclination of Albion to evolve on long timescales.

We see, then, that as the resonant angle $\phi_{TNO} - \phi_{Nep}$ approaches $0^{\circ}$, the forced terms are reductive, reducing the eccentricity and inclination of the TNO, and then as the resonant angle passes $0^{\circ}$, the forced terms become constructive, as Neptune applies a dragging torque to the small body's periapse over time.
This mechanism acts on the orbits of all of the low-inclination Classical belt TNOs, but only acts as the dominant frequency in the eccentricity and inclination evolution for TNO's that approach the resonant center of $g+s\approx g_8+s_8$.
 
In conclusion, we show that the $\nu_8+\nu_{18}$ secular resonance is in fact significant to the orbits of objects in the outer Solar System.
Because the resonance acts on eccentricity and inclination through gradual angular momentum exchange with Neptune's $g_8$ and $s_8$ secular modes rather than through any single close encounter, its influence builds up over timescales of hundreds of millions to billions of years, and can result in the some of the longest timescale instabilities in the Solar System.

\section{Understanding Classical Belt Inclination Architecture Through a Secular Lens}\label{sec:secular_architecture}

Having performed an analysis of the secular structure of the Classical belt TNOs, we have proven that the boundary of the Cold Classical belt TNO is strongly supported by an instability caused by the $\nu_8+\nu_{18}$ resonance.
The role of this resonance is to both destabilize many particles, while also lifting existing particles higher in inclination, suggesting that a subset of the currently Warm Classical belt TNOs may have originated at lower inclinations near the observed Cold Classical belt boundary. 
We discuss the impact of this realization in the following section, particularly with respect to measuring the initial population density of TNOs in inclination space.

\subsection{Classical Belt Inclinations}
\label{warm_class_sec}

The inclination distribution of the Classical Kuiper Belt has been the subject of considerable study, with a number of works making sustained efforts to more rigorously define the boundaries between the Cold and Hot Classical TNO populations \citep[e.g.][]{Brown:2001,Bernstein:2004, Volk:2011, VanLaerhoven:2019, Huang:2022}, hereafter referred to as CCTNOs and HCTNOs, respectively.
From the earliest surveys, a striking correlation was noted between dynamical class and surface properties: lower-inclination CCTNOs are generally redder and intrinsically fainter, while higher-inclination HCTNOs tend to be brighter and bluer \citep[e.g.][]{Tegler:2000, Levison:2001, Brown:2001}. 
This color–inclination dichotomy has led to the widespread use of dynamical classification as a proxy for surface composition and, by extension, formation environment, with the understanding that the true distributions are likely more complex \cite[e.g.][]{Bernardinelli:2025}. 
Reliably separating the CCTNO and HCTNO populations is therefore an important prerequisite for any population study of the outer Solar System.

A conservative inclination cut of proper $I_{free} < 4^\circ$ was identified by \cite{VanLaerhoven:2019} and subsequently confirmed by \cite{Huang:2022} as an effective means of isolating the CCTNO population, with very few HCTNO interlopers; those that do appear stand out clearly by virtue of their bluer colors \cite[e.g.][]{Fraser:2023}. 
Further supporting this boundary, \cite{Huang:2022} identified an apparent gap in the proper inclination distribution between the 5:3 and 7:4 mean-motion resonances, beginning at $I_{free} \approx 4^\circ$, which lends additional dynamical justification to the use of this cut.
%The upper boundary of the HCTNO population is considerably less well-constrained. 
CCTNO interlopers, identified via spectral analysis, are commonly found above the $I_{free} = 4^\circ$ boundary, likely having been displaced to higher inclinations by dynamical instabilities or sweeping mean-motion resonances during the final stages of Neptune's migration.
%As a result, there is a zone of significant mixing between the redder CCTNO and bluer HCTNO populations at intermediate inclinations. 

Studies aiming to isolate a clean HCTNO sample have consequently adopted higher inclination cuts of $I > 9^\circ - 12^\circ$, where contamination by red CCTNO interlopers becomes minimal \citep[e.g.][]{Peixinho:2008, VanLaerhoven:2019}.
\cite{Huang:2022} similarly notes that $I_{free} > 10^\circ$ effectively excludes CCTNOs, but stops short of characterizing the mixed population residing in the intermediate range $I_{free} < 10^\circ$, a region we refer to here as the ``Warm'' Classical Belt,  or WCTNOs, which has been observed to be the most dynamically and compositionally mixed region of the Classical Belt \citep{Peixinho:2008, Volk:2011, Gladman:2021}. 

%Recent color analyses confirm that the color-inclination distribution is likely much more mixed, especiallly among the Warm Classical belt. 

This is supported heavily by recent analyses of the surface color distribution of TNOs among the Classical belt.
\cite{Fraser:2023} recently demonstrated that TNO surfaces can be more accurately partitioned into two distinct taxonomic classes based on their reflectance in the near infrared. 
Building on this framework, \cite{Bernardinelli:2025} employed a Gaussian mixture model informed by Dark Energy Survey (DES) color measurements to quantify their distributions; they refer to the two groups as the Near-Infrared Faint (NIRF) TNOs, associated with the redder, more compositionally primitive objects, and the Near-Infrared Bright (NIRB) TNOs, corresponding to bluer objects thought to have been dynamically emplaced into the Classical Belt from the inner Solar System. 
Like the OSSOS survey, the DES survey was well-characterized in its on-sky selection criteria, allowing for an accurate debiasing of the inclination distribution measured by the survey.
Notably, \cite{Bernardinelli:2025} found the inclination distributions of NIRF and NIRB TNOs to be more complex than the simple bimodal picture would suggest; in particular, it would seem there exists a far more substantial population of NIRF objects appearing at higher inclinations than previously thought.
In other words, the color-inclination distribution of TNOs was found to be much more well-mixed at inclinations below $20^{\circ}$, indicating some mechanism had likely lifted many NIRF particles in inclination while largely preserving heliocentric distance.

In addition, there also exists a small subset of very low-inclination NIRBs found in the ``blue binaries'' \cite{Fraser:2017}. 
Taken together, these results indicate that meaningful mixing of NIRF and NIRB populations occurs across all inclinations, and that the classical color–inclination dichotomy, while a useful first approximation, does not cleanly separate the two formation populations.
In the absence of direct color measurements, population studies of the Classical Belt must therefore account for this mixing by employing model distributions that allow for contributions from both the CCTNO and HCTNO populations at all inclinations.

Our results from Section~\ref{sec:sec_mapping} offer a natural secular dynamical explanation for at least some of this mixing. 
Particles with initial inclinations near the boundary of the CCTNO belt are subject to interactions with the $\nu_8 + \nu_{18}$ secular resonance, which either lifts their inclinations or removes them from the Classical Belt entirely through Neptune scattering. 
This instability somewhat complicates the dynamical origins of the Warm Classical TNO (WCTNO) population, and partially accounts for why this region is so thoroughly mixed in color.
It also could explain the sharpness of the CCTNO truncation at $I_{free} = 4^\circ$: while a significant primordial population of CCTNOs at $4^\circ < I < 6^\circ$ is unlikely, a continuous inclination distribution almost certainly existed and could have been subsequently eroded by this mechanism.
These dynamical processes bear directly on efforts to reconstruct the primordial inclination structure of the belt. 

\cite{VanLaerhoven:2019} measured the mean plane and inclination distribution of the Cold Classical Belt from the OSSOS dataset \citep{Bannister:2018}, finding a measured inclination width of $\sigma_{CC} \approx 1.75^\circ$ relative to the mean plane predicted by Laplace-Lagrange secular theory.
This result has been interpreted as a constraint on the primordial Cold Classical Belt width, under the assumption that the observed and primordial distributions are similar. 
However, the selective removal of higher-free-inclination objects by the $\nu_8 + \nu_{18}$ resonance, along with upward inclination forcing from the Laplacian plane, means that the present-day observed distribution is likely a dynamically sculpted remnant of whatever primordial structure existed. 
Any reconstruction of the primordial belt must therefore deconvolve these effects, which we discuss in the next section.

\subsection{Filling the Gap: Measuring the Initial TNO Inclinations }
\label{sec:fitting_results}

Having identified a mechanism existing to remove particles from the $4^{\circ}<I_{free}<6^{\circ}$ region, we can posit two questions about the primordial TNO populations: 
(1) was the Cold Classical TNO belt originally much wider? and/or (2) did the Hot Classical belt initially extend to lower inclinations?\footnote{We can consider a third question which results from these first two: Did the particles which currently reside in the Warm Classical belt region form generally near their present day heliocentric distances, or were they primarily implanted from smaller heliocentric distances? This question gets to the heart of separating the primordially Cold Classical objects from the implanted Hot Classical objects.}
Since the $\nu_8+\nu_{18}$ secular resonance lies at the upper boundary of the Cold Classical belt and the lower boundary of the Hot Classical belt, this mechanism appears to only deplete the tails of those two components located within $4^{\circ}<I_{free}<6^{\circ}$, leaving the cores of the Hot and Cold components minimally changed. 

From our grid of simulated particles over 4.5 Gyr, we can measure the expected final shape of the proper inclination distribution of both the Cold and Hot Classical belt TNOs given hypothesized initial inclination distributions and make comparisons to the observed populations.

To do this, we compute a transfer function to map initial orbital elements to their 4.5 Gyr evolved states which is constructed directly from the grid of simulated particles described in Section \ref{sec:sec_mapping}.

This transfer function reflects purely dynamical information, and can be considered the mapping of which initial conditions are dynamically accessible given the observed proper element, weighted by the density of stable trajectories connecting them.
We can use the transfer function to solve for the best-fit initial inclination distribution parameters of a mixed-distribution model which contains a Cold Classical component and a Hot Classical component based on a comparison with the observed TNO populations.

It's important to note that because many wide-field surveys concentrate their pointings near the ecliptic, they preferentially sample the region of sky where low-inclination TNOs spend nearly all of their time. 
High-inclination objects, by contrast, spread their time over a much larger range of ecliptic latitudes as they move through their orbits, so any fixed on-plane survey footprint captures only a small, latitude-dependent fraction of the high-inclination population.
In other words, there exists significant bias in the catalog of known TNOs, favoring detections near the ecliptic due to the on-sky selection bias of surveys at low-ecliptic longitudes (see, e.g., \citealt{Kavelaars:2020} for a review of outer solar system observational biases).
Correcting for this effect requires detailed knowledge of each survey's pointing history and detection efficiency as a function of ecliptic latitude, which requires debiasing  that is available with survey simulators only for a small subset of well-characterized surveys (see, e.g., \citealt{Lawler:2018}).

To mitigate the inclination bias in  our results, we restrict our present analysis to only include objects with $I<10^{\circ}$.
%Within this low inclination range, essentially the full range of classical belt orbits is detectable by typical ecliptic-survey footprints regardless of pointing strategy, so the observed sample can be treated as effectively unbiased without requiring survey-specific debiasing.
Such low-inclination classical belt orbits typically fall within the latitude range of ecliptic-focused surveys, allowing us to avoid the need for survey-specific debiasing.
This lets us draw on the much larger set of TNOs discovered across all surveys, rather than restricting ourselves to a smaller sample from one or two well-characterized surveys.

We fit both components to a von Mises-Fisher distribution, which has been shown in previous work to best represent the distribution of orbital orientations \citep{Matheson:2023}.
The von Mises-Fisher distribution is particularly useful, as it allows for the measurement of an inclination center on the surface of a unit sphere, $\boldsymbol{\hat\mu}$, as well as a $\kappa$ parameter which measures the concentration of the inclinations of the sample near the center.
The location of the center of the distribution, $\boldsymbol{\hat\mu}=[0,0]$, is defined by the choice of reference plane, which in the case of our proper element catalog of osculating elements is defined as the invariable plane.
However, we note that as the proper elements are defined with respect to the forced plane of the outer Solar System, directly fitting the proper element distribution to a von Mises-Fisher distribution would instead place the center $\boldsymbol{\hat\mu}=[0,0]$ at the location of the locally forced plane.
In our analysis, we marginalize over the longitude variable $\phi$ by assuming the center is always aligned at $\phi=0^{\circ}$, allowing us to reduce the von Mises-Fisher distribution to a 2-dimensional distribution on the spherical surface, corresponding only to the inclination of the center of the distribution, defined as the now 1-dimensional $\mu$, which is a measure of inclination relative to the invariable plane at $\mu=0^{\circ}$, and the concentration parameter, $\kappa$.

When the inclinations are strongly concentrated around the center (i.e., $\kappa\gg1)$, the inclination distribution function simplifies to a Rayleigh distribution, with the relation between the Rayleigh width parameter $\sigma$ and the $\kappa$ parameter being $\sigma=1/\sqrt{\kappa}$; thus there is a direct relation which can represent the concentration of a von Mises-Fisher distribution as an inclination width, $\sigma$, which can be expressed in units of degrees. 
A full description of our fitting methodology are discussed in detail in Appendix \ref{app:fitting_incs}. 

The result of our fitting for a primordial cold population generally finds a best-fit concentration parameter of $\kappa_{CC} \approx 1455$ with a center lying along the invariable plane, corresponding to $\sigma_{CC}\approx1.5^{\circ}$ as defined by the Rayleigh distribution. 
In contrast, we find that the hot population component fits are largely unconstrained, due to our limited range of considered inclinations at $I_{free}<10^{\circ}$, with a wide range of non-zero centers and varying component widths all fitting generally well to the data.
Within the range of inclinations we consider, the fraction of particles belonging to the Cold Classical component $x\approx 0.88$, with the remainder of the  
This result is generally consistent among multiple selected semi-major axis bins, which are highlighted in Appendix-\ref{app:fitting_incs}, demonstrating the most likely Cold Classical width overall, though the mixture fraction of the Cold Classical component shrinks at higher semi-major axis.

\begin{figure*}[htb]
    \centering
    \includegraphics[width=1.0\linewidth]{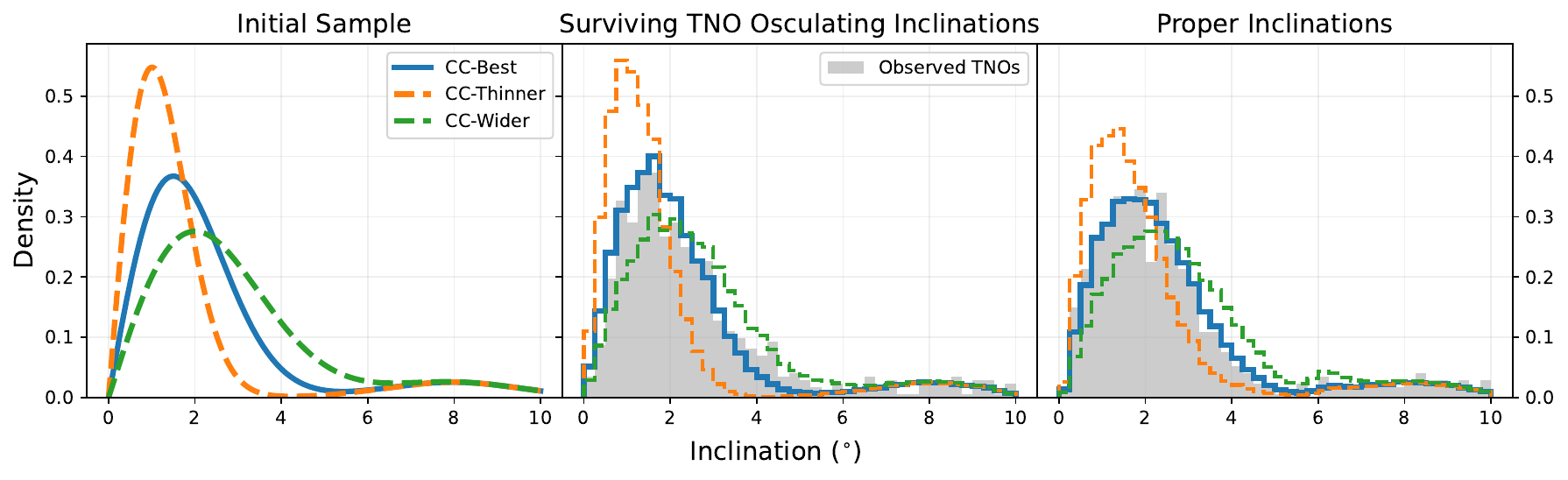}
    \caption{Sampled initial inclination distributions and resulting stable distributions for several mixed von Mises-Fisher inclination distributions applied to the Cold Classical belt region from $42.3<a<44.5$ au at $I_{free}<10^{\circ}$ (left panel). The CC-Best model (blue) is generated from the best-fit parameters measured by our analysis, which has $\mu_{CC}\approx0^{\circ},\; \sigma_{CC}\approx 1.5^{\circ}\; \mu_{HC}=7.9^{\circ},\; \sigma_{1.55^{\circ}}$. The CC-Thinner (orange) and CC-Wider (green) models maintain all of the same parameters as the CC-Best model, but have modified $\sigma_{CC}=1, 2^{\circ}$, respectively. 
    The hot component of the mixed distribution is held constant for all three models.
    Samples matching these initial distributions are drawn from our grid of particles, and the corresponding surviving particles after 4.5 Gyr of evolution are shown in the center (osculating inclination distribution) and right (proper inclination distribution) panels. 
    The observed TNO sample for the same semimajor axis and proper inclination range is shown in the right two panels are shown in grey.
    Instability notably impacts the shape of the final distribution in both osculating and proper inclination space in measurable ways.
    \label{fig:comp_fits}}
\end{figure*}

As the transfer function represents a transformation from an initial distribution of particles to the resulting proper elements, our best-fit model is representative of an initial distribution of particles as defined in osculating element space, which have a one-to-one transformation to some proper element distribution after 4.5 Gyr of dynamical sculpting and evolution.
This is visualized in Figure~\ref{fig:comp_fits}, which demonstrates in three separate panels (1) the initial sample of TNOs drawn from the best-fit inclination model, and is representative of a mixed von Mises-Fisher distribution, (2) the distribution of stable particles in osculating inclination space fro the drawn sample, and (3) the resulting proper element distribution of the remaining stable particles. 
Figure~\ref{fig:comp_fits} specifically shows the observed catalog of TNOs and best-fit model corresponding to the region between the 5:3 and 7:4 mean motion resonances, which region is most strongly impacted by the $\nu_8+\nu_{18}$ resonance discussed in Section-\ref{sec:gs8_res}.
In addition, this figure also includes the model with a wider and a narrower primordial width for the Cold Classicals ($\sigma_{CC}$).
We note that while the 1st panel of Figure~\ref{fig:comp_fits} clearly demonstrate smooth mixed von Mises-Fisher distributions, dynamical sculpting causes the resulting population in both osculating and proper space to no longer represent a von Mises-Fisher function.

It's clear that an initial width of $1.5^{\circ}$ does represent a better fit to the observed TNO catalog of proper inclinations than a wider primordial Cold Classical component, though there remains some observational noise within the observed TNO population which prevents total agreement, especially in the 2nd panel.
However, it is notable how well the best-fit model appears to fit to the observed proper inclination distribution, despite demonstrating some notable disparity in the 2nd panel.
This comparison highlights the feasibility of our method, which doesn't fit directly to the osculating elements themselves, but instead marginalizes over the probability space of the initial osculating elements which best produce the observed proper elements today. 

We mention also an interesting feature which results in the proper element distribution of the wider model.
The model which implements a $\sigma=2^{\circ}$ includes a large number of particles with osculating $4^{\circ}<I<6^{\circ}$.
Due to interactions with the $\nu_8+\nu_{18}$ resonance, instability combined with resonance sticking at the resonant center causes a ``bump'' to appear near $I_{free}\approx6^{\circ}$.
The resonant bump appears to be consistent feature which appears in any model that includes significant population of particles initially within this region.
While the best-fit model to the presently observed catalog indicates this region was likely initially sparse, we point out that there does appear to be a small ``bump'' in the observed catalog at $I_{free}\approx 6^{\circ}$. 

The most recent measurement of the Cold Classical belt width was done by \cite{VanLaerhoven:2019}, who reported a best-fit present-day Cold Classical belt width of $\sigma=1.75^{\circ}$ as contained in the OSSOS dataset, across multiple semi-major axis bins.
Their study also indicated a slight rejection of $\sigma=1.5^{\circ}$ for 2 of the 3 ranges of semi-major axis that they sampled in their analysis, presenting a slight discrepancy between their result and our fits.
However, this discrepancy between our results and the results of \cite{VanLaerhoven:2019} likely occurs due to (1) a difference in the sample size we use and the OSSOS dataset, and (2) differences in the actual method of measurement of the Cold Classical belt width. 
In the first case, \cite{VanLaerhoven:2019} mentions that in their samples for the inner-main, kernel, and outer-main regions, the reported sample sizes for their analyses are 107, 82, and 67 nonresonant Cold Classical TNOs at $I_{free} < 4^{\circ}$, respectively.
In contrast, when considering the same semi-major axis ranges, and considering only TNOs at $I_{free} < 4^{\circ}$, our sample includes 293, 324, and 199 nonresonant CCTNOs; however, as our analysis solves for the mixed 2-component model, we also include TNOs from $4^{\circ}<I_{free}<10^{\circ}$, which increases our sample.
When including these additional TNOs, our full sample size we implement for each of these semi-major axis ranges increases to 353, 357, 358 TNOs.

In the second case, while \cite{VanLaerhoven:2019} directly measured the width of the \textit{current} proper inclination distribution of their dataset, we measure the probable \textit{initial} width in osculating element space that, once convolved with 4.5 Gyr of secular evolution, results in a specific proper inclination distribution fit to today's population; this final proper distribution could be skewed towards slightly higher inclinations than the initial one due to secular lifting and destabilization.
Despite these differences between the studies, both results lie within reasonable distance of each other, and when we apply our analysis to the OSSOS dataset, we retrieve very similar results to those corresponding to this full dataset.
We provide some additional discussion and comparison of the two studies in Appendix~\ref{app:fitting_incs}.

We favor our result due our larger sample size, the visual correspondence between the best-fit sample in proper element space and the observed catalog, and the physical interpretation of our result with respect to the primordial disk.
However, we note that our result should improve significantly with an increased catalog size, and better characterization of the on-sky selection rate by future surveys like LSST.

In any case, our result agrees strongly with the conclusion made by \cite{VanLaerhoven:2019}, which is that Neptune's migration likely left the Cold Classical particles below $a<45$ au fairly undisturbed, highlighting the primordial nature of this region of Trans-Neptunian space, and the resulting particles within it.

We also report that while the two populations are still likely best described by a mixed component model, it is likely that there is very little contamination of implanted Hot Classical particles mixed into the higher inclination Cold Classical region around $I\approx 4^{\circ}$.
While our ability to constrain the real shape of the Hot Classical belt is limited by the truncation in inclination we implement, none of our results indicate a Hot Classical component which contributes significantly to the $I_{free}<4^{\circ}$ region of TNOs.
In fact, most of our models favor placing the center of the Hot Classical component above $0^{\circ}$, seemingly in an effort to keep the $4^{\circ}<I_{free}<6^{\circ}$ region mostly depopulated.
We also point out that the mechanism of the $\nu_8+\nu_{18}$ acts primarily to lift inclinations; thus it is highly unlikely that secular forcing could cause Hot Classical particles to evolve to lower inclinations, further reinforcing the idea that Hot Classical particles should be largely absent from the Classical belt at $I_{free}<4^{\circ}$. 

This method of course relies directly on the assumption that all of these particles experience no additional effects beyond simple interactions with the giant planets over time, which is not necessarily true.
Other effects may play a role in the secular evolution of particles, especially at early times within the Solar System history; these include the presence of additional planets which have since been ejected from the Solar System \citep[e.g.][]{Gladman:2006,Nesvorny:2012}, self-gravity of the massive disk \citep[e.g.][]{Kaib:2024}, collisional or gravitational interactions with other massive TNOs \citep[e.g.][]{Abedin:2021,Munoz-Gutierrez:2019}, or binary TNO interactions which have larger cross-sections \citep[e.g.][]{Campbell:2023} and could have been much more common in the early Solar System if the primordial was much larger, as has been suggested by the streaming instability model \citep[e.g.][]{Youdin:2005, Nesvorny:2019}. 
In addition, this analysis makes no attempt to measure or account for the impact of sweeping mean motion or secular resonances during planetary migration on the inclination distribution of the particles; we simply measure the impact of secular diffusion and inclination lifting from the present-day known Solar System quantities. 
A future study looking at the impact of mean motion resonant sweeping in context of present migration models on the inclination distribution of the Classical belt particles will be considered. 

Our final result indicates that the region of the Classical belt from $4^{\circ}<I_{free} < 6^{\circ}$, was most likely the least populated region of space within the Classical belt post-Neptune migration.
In other words, while the $\nu_8+\nu_{18}$ resonance produces a region of instability within the Classical belt, there were likely very few particles to actually displace, or the mechanism removed more particles than current estimates would expect.
%Does this mean that Neptune's present day orbit was perhaps guided in some way 
Does this mean that Neptune's orbit evolved such that its angular momentum vector which minimized the amount of angular momentum exchange with the TNOs in the outer Solar System? 
Or could the $\nu_8+\nu_{18}$ resonance have been at some point in Neptune's history more severe, if Neptune had a larger eccentricity and/or inclination, which were damped over time as these very particles were scattered by Neptune?
While beyond the scope of this study, we expect the increased LSST catalog to make clear which answer may be the most likely.

\section{Conclusions}

In this paper, we compute proper orbital elements for all known TNOs using \texttt{SBDynT} and present them in a machine-readable table for community use, accompanied by uncertainties derived from the proper element computation. 
To contextualize these proper elements within the mean dynamical behavior of the outer Solar System, we also integrate and compute proper orbital elements for a synthetic sample of 400,000 Classical Belt TNO particles spanning $40<a<48$ au. 
The resulting secular maps of the Classical Belt reveal that the long-term $\nu_8 + \nu_{18}$ secular resonance, with a circulation period on the order of $66$ Myr, plays a significant role in depopulating TNOs with $4^\circ < I_{free} < 6^\circ$, demarcating the dynamical boundary of the Cold Classical Belt.

Our mapping of this instability at the Cold Classical Belt boundary motivates a careful estimation of the primordial belt's inclination width, accounting for secular effects that lift particle inclinations on gigayear timescales while conserving semi-major axis. 
By fitting a mixture of two von Mises-Fisher distributions to a probabilistic transfer function that maps proper elements back to initial conditions after 4.5 Gyr of dynamical evolution, we find that the primordial Cold Classical Belt is best described by a population centered on the invariable plane with a width of $\sigma_\mathrm{CC} \approx 1.5^\circ$, which is slightly thinner than previous estimates made by \cite{VanLaerhoven:2019} of $\sigma_{CC}\approx1.75^{\circ}$. 
This result reconfirms that the primordial belt was closely aligned with the present-day invariable plane, and is consistent with previous estimates of its strikingly small inclination width indicating that the Cold Classical belt TNOs have largely been left dynamically untouched since formation.

Several improvements to our results can be made as the LSST survey catalog comes into view in the next few years.
First, the well-characterized LSST survey catalog will allow us to account for bias at higher inclinations; thus we can fit to our model TNOs with $I_{free} > 10^\circ$ while avoiding the impact of bias introduced by the on-sky selection rate of observing the outer Solar System, which limits our current analysis strictly to measuring the Cold Classical belt inclination distribution.
Second, we do not incorporate surface color taxonomy into the fitting in the present work, but doing so could provide critical additional constraints on TNO origins and population mixing.

A follow-up analysis could use the measured color catalog of LSST TNOs to cleanly separate TNOs into the NIRF and NIRB classes, independently fit each population's inclination distribution with individual von Mises-Fisher models, and then compare the shapes of these individual color-based class components to the dynamical analysis we have done here.
Close agreement between the NIRF and NIRB color-based components with the measured dynamical Cold and Hot components in our analysis would indicate that secular evolution alone could plausibly explain the inclination distribution of the NIRF population, supporting the interpretation that the NIRF particles formed with a single primordial reservoir from $\approx40<a<48$ au, with a very narrow inclination width. 
Disagreement between the two models, on the other hand, could provide insight into the level of dynamical lifting needed to bridge the gap between secular forcing and the observed population, or whether the model of in situ-formation of the outer solar system TNOs requires modification.

We note that extending this analysis to higher inclinations will also require careful treatment of TNO collisional families. 
The Haumea family is already statistically prominent in the current catalog, a consequence of elevated albedos among family members and the inherently prolific nature of collisional disruption, and we find that even after removing confirmed family members, a statistically significant overdensity persists at $25^\circ < I_{free} < 29^\circ$, suggesting that a substantial number of unconfirmed Haumea family members remain in the known catalog. 
As the TNO catalog grows, the influence of collisional families on population-level analyses will only increase, potentially even revealing new collisional families which stand out among the inclination distribution of these regions.

The LSST era is poised to resolve many of these open questions. 
The dramatic expansion of the TNO catalog, combined with survey-characterized selection functions and near-universal color measurements, will enable a direct extension of this analysis, simultaneously constraining the primordial inclination distribution, the degree of secular sculpting, the color-inclination mixing fraction, and the census of collisional families.

\vspace{12pt}
\noindent{\it Acknowledgments:}
This work was supported by NASA grants 80NSSC23K0886 (PDART), and 80NSSC23K1374 (FINESST). 
KV acknowledges additional support from NASA grants 80NSSC23K0680, 80NSSC26K1026, and 80NSSC26K1031.

%%%%%%%%%%%%%%%%%%%%%%%%%%%%%%%%%%%%%%%%%%%%%%%%%%%%%%%%%%%%%%%%%%%%%%%%%%%%%%%%%%%%%%%%%%%%%%%%
\facilities{ADS}
\software{rebound, SBDynT}
%%%%%%%%%%%%%%%%%%%%%%%%%%%%%%%%%%%%%%%%%%%%%%%%%%%

\bibliography{ADSrefs, customrefs}
\bibliographystyle{aasjournalv7}

\newpage
\begin{appendices}

\section{Measuring the Transfer Function from Primordial to Proper Distributions}
\label{app:fitting_incs}

With the realization that the proper inclinations preserve some memory of the initial distribution of particles within the post-migration architecture, we develop a Bayesian framework to infer the primordial inclination distribution of the Classical Kuiper Belt, accounting explicitly for the dynamical mapping between initial orbits 4.5 Gyr ago and current orbital elements.
Rather than fitting directly to observed osculating or proper inclinations, we construct a probabilistic transfer function that connects each observed TNO to the distribution of initial conditions from our large grid of integrated particles (Section~\ref{s:secular_map}) consistent with its current dynamical state.

For each observed TNO $k$ in our catalog with proper elements $(a_{p,k},e_{p,k}, I_{p,k})$, we identify the 5 nearest neighbors in the three-dimensional proper element space $(a_p,e_p,\sin{I}_p)$ among the simulated grid of particles using a KDTree with a Euclidean metric; Euclidean distances in proper element space can be approximated using the distance metric, $D_k = \left(\sqrt{5/4}\,a_{k}/\langle a\rangle, \sqrt2 \,e_{k}, \sqrt2 \, \sin{I_{k}}\right)$. 

We note that while this grid is well-sampled enough throughout parameter space to allow for accurate measurement of the mean secular behavior of particles within the specific grid shown in Section \ref{sec:sec_mapping}, our transfer function requires each region to be highly sampled for accurate measurement of the primordial disk width.
In particular, the low-a particles among the Cold Classical belt TNOs are somewhat depopulated, as stable orbits only exist for low-e particles.
To better sample the Cold Classical population, we integrate and compute proper elements for an additional 100,000 particles at $a>42$ au at $I_{init}<10^{\circ}$.

Each neighbor $j$ is assigned a weight $w_{k,j}$
%w_kj = exp(−½ (d_kj / h)²) / Σⱼ exp(−½ (d_kj / h)²
\begin{equation}
    %w_{i,p} = \exp\left[-\frac{1}{2}(D_{kj}/h)^2\right] / \sum_j \exp(-\frac{1}{2}(D_{kj}/h)^2)
    w_{kj} = \frac{\exp\left(-\frac{1}{2}\left(\frac{D_{kj}}{h}\right)^2\right)}{\sum_{j} \exp\left(-\frac{1}{2}\left(\frac{D_{kj}}{h}\right)^2\right)}
    \label{weights}
\end{equation}
where $D_{kj}$ is the distance in $(a_{p},e_{p}, \sin I_{p})$, and $h$ is an adaptive bandwidth set to the median neighbor distance for that TNO.
The weighting is performed entirely in proper element space; the initial inclinations $I_{0,kj}$ of the neighbors are carried passively as labels and are never used in the distance computation. 

We then evaluate a mixed von Mises-Fisher distribution over the probabilistic initial conditions represented by the transfer function, which distribution has been shown to most accurately represent the distribution of small body inclinations \citep{Matheson:2023}.
This construction naturally handles the multi-modal structure introduced by secular resonances and instability, which may cause objects originating at different initial inclinations to converge at the same proper element through separate dynamical pathways.
%All transfer weights are precomputed once prior to the MCMC sampling, as they are independent of the model parameters, reducing the likelihood evaluation to a fully vectorized operation.

The Classical belt TNOs have multiple distinct bins of semi-major axis, but we fit inclination distributions to three specific semi-major axis ranges, (1) The inner-main portion of the main belt between the 5:3 and 7:4 mean motion resonances ($42.3 < a < 43.7$ au),  (2) The ``kernel'' region between the 7:4 and 9:5 resonances ($43.7<a<44.5$ au)\footnote{We point out that while the kernel typically refers to the cluster of Cold Classical particles at $I_{free}<4^{\circ}$ that lie between the 7:4 and 9:5 mean motion resonances, we use the term in this section to refer to all of the particles used in our analysis which lie between those mean motion resonances. The kernel simply acts as a guardrail for this range of semi-major axes.}, and (3) the outer-main portion of the main belt between the 9:5 and 2:1 resonances ($44.5<a<47.8$ au).
We also fit the inclination distribution of all three semi-major axis ranges at once.
We do not include the $40<a<42$ au range of particles, as the $\nu_{18}$ resonance effectively wipes out any low-inclination particles, leaving us with no Cold Classical component to fit to.

\begin{equation}
    vMF(I \mid \mu, \kappa) = \exp\left[\kappa\cos(I - \mu) - \kappa - \ln (2\pi\,I_0^e(\kappa))\right]
    \label{eq:vonmises}
\end{equation}
\\

We fit the primordial inclination distribution of these regions with 2-component von Mises-Fisher distributions, representing a Cold Classical population and a Warm Classical population.
The von Mises-Fisher distribution has been shown to be the best representation of small body orbital orientations in orbital phase space \cite{Matheson:2023}, and is thus preferred in our study of the inclination distribution.
In the formulation of the von Mises-Fisher probability density function shown in Equation \ref{eq:vonmises}, the parameter $\mu$ represents the center of the distribution, and the parameter $\kappa$ represents the concentration of the distribution.
For values of $\kappa<<1$, the distribution approaches a uniform distribution at every angle, while at large $\kappa$, the distribution approaches a normal distribution, with $\kappa\approx1/\sigma^2$.
In this case, $I^e_0(\kappa)=e^{-\kappa}I_0(\kappa)$ is the exponentially scaled modified Bessel function of the first kind.
We adopt the von Mises-Fisher distribution in preference to the times $\sin I$ form first employed by \citet{Brown:2001}, which artificially suppresses the population near $I=0^{\circ}$ through the $\sin I$ Jacobian term. 
% This suppression is appropriate for the current observed population, where secular forcing by Neptune has had time to phase-mix the distribution, and push particles generally above the Laplacian plane at lower semi-major axes, but is physically unwarranted for a primordial disk prior to the emplacement of Neptune, where a large population of near-coplanar objects is entirely plausible. 
% The von-Mises distribution evaluated near $\mu=0^{\circ}$ recovers half-Gaussian behavior without imposing the unphysical zero at $I=0^{\circ}$, while its concentration parameter $\kappa$ smoothly interpolates between a near-uniform distribution (small $\kappa$) and an arbitrarily cold spike (large $\kappa$).
% For numerical stability at large $\kappa$, we evaluate the PDF entirely in log space using the exponentially scaled Bessel function, avoiding the overflow that would otherwise occur near $\kappa\approx720$ when computing $I_0^{\kappa}$ directly in double precision.

The log-likelihood of a fitted mixed model distribution is constructed by marginalizing over the transfer function for each observed object.
We additionally weight the transfer function by the stability of the initial osculating elements of the resulting nearest neighbors; while secular resonance and mean motion resonance can cause particles to migrate to new regions of phase space, this process is often chaotic, and the resulting orbits is very sensitive the the initial conditions. 
For this reason, we want to suppress the weight of neighbors which may originate near less table regions of parameter space, which allows us to account for resonance instabilities in our analysis.

We demonstrated several measures of stability in Section \ref{sec:sec_mapping}, but the simplest measure of stability in a region of osculating phase space would be to simply measure how many particles within some region of osculating phase space remain within the Classical belt by the end of the integration. 
To do this, for each particle, we recorded its final osculating elements and flagged it as "surviving" if it remained bound to the classical belt region, defined as $40 < a < 50$ au and $e<1$, at the end of the integration. Counting the number of surviving particles in each initial $(a_0, e_0, I_0)$ bin and normalizing by the total number of particles initially placed in that bin yields a survival fraction $\mathcal{S}_0(a,e,I) \in [0,1]$, representing the empirical probability that a particle with those initial osculating elements remains dynamically stable over the age of the Solar System. 

To evaluate the survival fraction at arbitrary $(a_0,e_0,I_0)$ not coincident with a grid node, we linearly interpolate over this discrete grid using scipy's  \textbf{RegularGridInterpolator}, allowing $\mathcal{S}$ to be incorporated as a continuous weighting function within our probabilistic likelihood framework.
Using a grid with bins of $\Delta a = 0.2$ au, $\Delta e=0.01$, and $\Delta I = 0.5^{\circ}$, we confirm that the median number of particles per bin in the initial grid is approximately 15, indicating that our grid resolution is sufficient for some measurement of stability.

For an observed TNO, $k$, the per-object likelihood contribution for each component $i$ within the mixed model is given by

\begin{equation}
    c_{k,i} = x_i \cdot \frac{\sum_j w_{kj} \cdot vMF_i(I_{kj}) \cdot \mathcal{S}(a_{kj}, e_{kj}, I_{kj})}{\mathcal{Z}_i} 
    \label{2mixed}
\end{equation}
where the normalization constants $\mathcal{Z}_{i}$ are computed analytically by evaluating the von Mises-Fisher PDF over a fine grid of inclinations spanning $[0^{\circ}, 180^{\circ}]$, independently of the simulation particle grid as
\begin{equation}
    \mathcal{Z}_{i}(\mu, \kappa) = \frac{1}{N}\sum_{\theta=1}^{N} vMF_i\!\left(I_{0,\theta} \mid \mu, \kappa\right), \quad I_{0,\theta} \in \left[0^{\circ},\, 180^{\circ}\right]
    \label{norm}
\end{equation}
$x_i$ represents the the mixture weight of each component in the model, subject to the simplex constraints $x_i>0,\,\;\; \sum_ix_i<1$. 

Posterior distributions are sampled using the affine-invariant ensemble MCMC sampler \texttt{emcee} \citep{Foreman-Mackey:2013}.
The best-fit model can be visually validated by sampling particles according to the best-fit model inclination distribution, and compared to the observed TNO inclination distribution.

\subsection{Synthetic Recovery Testing}

To assess the sensitivity and accuracy of our method of fitting a distribution to the probabilistic transfer function, we perform a suite of synthetic recovery tests using the simulation particle catalog itself. 
We partition the 200,000 simulation particles into two subsets: three-quarters are reserved to construct the KDTree transfer function, and the remaining quarter are used to generate synthetic observed catalogs with known primordial distributions.
For each test, we draw a synthetic observed sample from the quarter-catalog by weighting particles according to a specified initial two-component von Mises-Fisher mixture with known parameters $\boldsymbol{\theta}_\text{true}$, reading off the proper elements of the drawn particles at 4.5 Gyr as mock observations. 
The fitter is then applied to these mock observations using the three-quarter transfer function catalog, and the recovered posterior is compared to $\boldsymbol{\theta}_\text{true}$.

We find that the framework reliably recovers the Cold Classical inclination center $\mu_\text{CC}$ to within approximately $0.1^{\circ}$, with the true value falling within the $1\sigma$ credible interval in all tested configurations.
This sensitivity is sufficient to resolve the expected range of primordial cold disk inclinations and to distinguish meaningfully between a disk centered precisely on the invariable plane and one with a small but nonzero primordial tilt.

However, the recovery tests reveal a systematic degeneracy near $\mu_\text{CC} = 0^{\circ}$.
When the true cold classical center lies at or very close to zero, the posterior admits solutions in which a slightly broader CC component centered at $\mu_\text{CC} \sim 0.1-0.15^{\circ}$ provides a comparably good fit, compensated by a correspondingly different HC component. 
This degeneracy arises because objects with initial inclinations near $I_0 \sim 0^{\circ}$ are efficiently lifted to the Laplacian plane by secular forcing, producing a proper element distribution that is difficult to distinguish from a slightly warmer primordial distribution without forced lifting. 

As a result, the individual component parameters become partially degenerate near $\mu_\text{CC} = 0^{\circ}$, though the mixture as a whole remains a good description of the proper element distribution.
Recovery of the hot classical component is less robust in general. 
Because the HC distribution is broad and overlaps significantly with the high-inclination tail of the CC and stirred populations, the fitter has a tendency to absorb excess CC population variance into the HC component, effectively overfitting the cold population at the expense of a well-constrained hot component.
We find that HC recovery improves substantially when $\mu_\text{CC} \gtrsim 1^{\circ}$, in which case the two populations are sufficiently separated in initial inclination space that the transfer function can more fully distinguish their contributions. 

These recovery tests establish that our primary scientific conclusions regarding $\mu_\text{CC}$ are robust at the $0.1^{\circ}$ level, while the resulting best-fits regarding the HC component parameters should be interpreted with appropriate caution, particularly in cases where the CC population is consistent with $\mu_\text{CC} \approx 0^{\circ}$.
We report HC parameters as simply indicative of the broad character of the hot population rather than precise measurements, and we note that a fully characterized survey simulator would be required to place stronger constraints on the hot classical inclination distribution from the heterogeneous compiled catalog.

\subsection{Model Fits}
\label{sec:model_fits}

\begin{figure*}[!ht]
    \centering
    \includegraphics[width=1\linewidth]{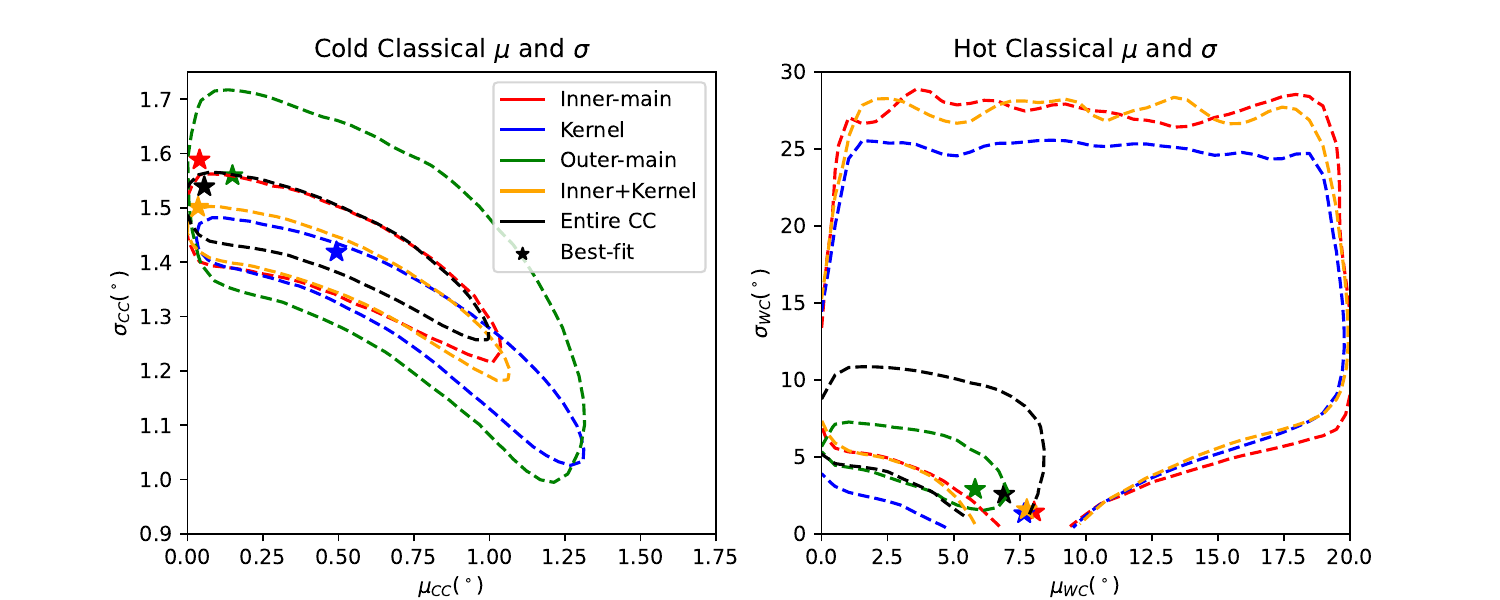}
    \caption{The $1\sigma$ contours for the distribution of primordial Cold Classical and Hot Classical inclination centers and widths for the transfer function computed with the $k=5$ nearest neighbors. The best-fit parameters are shown as stars. The width is reported in $\sigma$, computed according to $\sigma\approx 1/\sqrt{\kappa}$. The inner-main belt Cold classical component best-fit notably lies outside of the 1-sigma boundary, primarily since it lies so close to the edge at $\mu=0^{\circ}$, causing fewer samples to be taken near that space, despite producing high-likelihood fits. }
    \label{fig:cc_contours}
\end{figure*}

The 1-sigma posterior distributions for the fitted models are shown in Figure \ref{fig:cc_contours}. 
We report the best-fit widths in terms of $\sigma$-values, which are computed directly from the fitted $\kappa$ parameters by way of the approximation $\sigma\approx1/\sqrt{\kappa}$.
This is done so that the widths may be reported in a functional form that carries units of degrees and can be more easily compared to other works.

We emphasize that due to the inclination limit we place on the observed TNO distribution of only considering TNOs with $I_{free}<10^\circ$, the model fits produced for the Warm Classical component of the model is highly unlikely to capture the true dynamical behavior of these TNOs.
Indeed, we see in these Figures that for every region we test, the second component in the model is largely unconstrained, with centers and widths that span 10's of degrees in either direction.

However, this mixed model allows us to measure well the potential distributions for the Cold Classical belt TNOs.
To make direct comparison between the Cold Classical model fits, we show in Figure \ref{fig:cc_contours} the $1\sigma$ contours and the best-fit parameters for each regions posterior distribution.
The functional correlation between the inclination center and the inclination width is instantly noticeable. 
We point out that this degeneracy occurs because the fitter attempts to place the peak of the inclination distribution at approximately the same point, while shifting the shape and placement of the wings. 
Because of this, centers and widths along the same function curve can be considered as producing the same overall peak in the inclination distribution while modifying the wings on either side of that peak.
The placement of the best-fit parameters provide some visible guidance as to the true distribution; we see that a $\mu=0^{\circ}$ is preferred by every model fit, indicting that the post-migration belt was very likely aligned with the invariable plane. 

\begin{figure}
    \centering
    \includegraphics[width=1\linewidth]{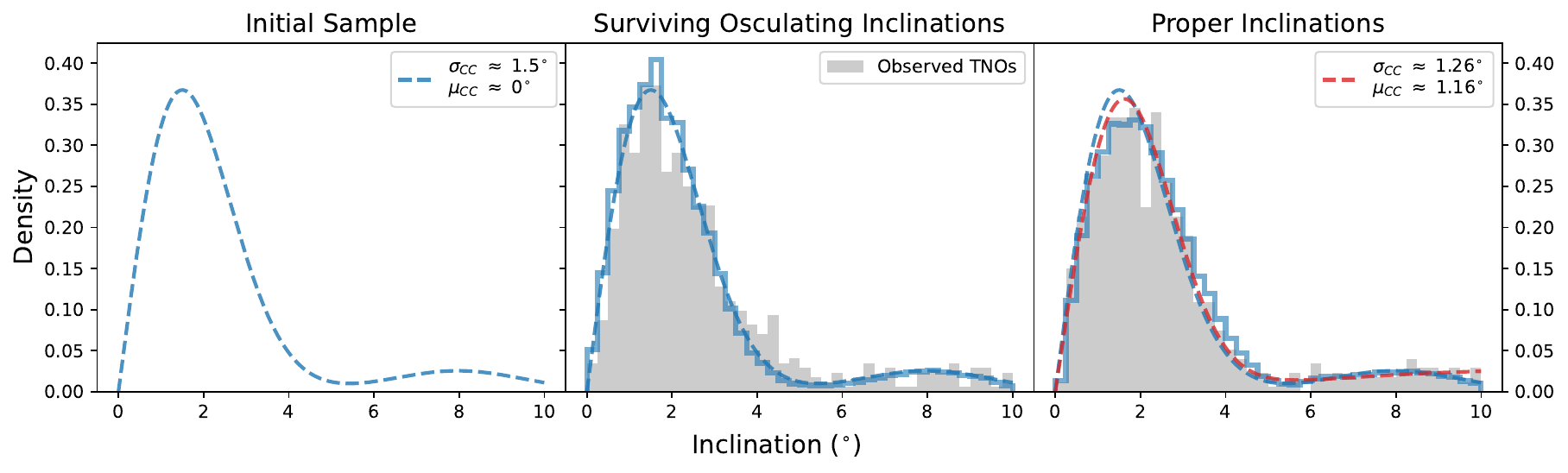}
    \caption{Similar to Figure-\ref{fig:comp_fits}, the best-fit model of the transfer function to the catalog of known TNOs in both the inner-main and kernel regions, a sample of TNOs drawn from that best-fit from our grid of synthetic particles, and the resulting proper element distribution of this sample after 4.5 Gyr. A direct fit of the mixed von Mises-Fisher model to the proper element distribution of the known TNOs is also shown in red. As the initial population is shaped by instability over 4.5 Gyr, the resulting proper inclination distribution is no longer well-represented a von Mises-Fisher distribution. Instead, the proper inclination distribution peak appears flatter and shifted slightly upwards from the initial von Mises-Fisher distribution.}
    \label{fig:comp_prop}
\end{figure}

In contrast, the inner-main belt and kernel Hot Classical model fits are generally unconstrained, as any function with a peak greater than our $10^{\circ}$ upper limit can generally be considered a good fit to the data. 
However, the fits to the outer-main belt and the entire Classical belt indicate a more well-constrained Hot Classical belt.
In both cases, a $\mu_{HC}=0^{\circ}$ is rejected, as is any width $\sigma\lesssim 5^{\circ}$.\
While these model fits are somewhat impacted by the upper limit placed on the observed inclinations at $10^{\circ}$, it is visibly clear in Figures \ref{fig:small_catalog} that there is indeed an overdensity of particles within the Warm Classical region of the outer-main belt, which is the only compoennt of the Hot Classical belt we sample in this analysis.  

The requirement that the center of the Hot Classical component be greater than $I_{free}=0^{\circ}$ is primarily resulting from the fact that any von Mises-Fisher distribution of Hot Classical belt TNOs centered at $I_{free}=0^{\circ}$ would cause the Hot Classical component to contribute a non-negligible number of TNOs to the low-inclination TNO region of $I_{free} < 6^{\circ}$.

We point out again that, when considering the impact of instability which selectively removes particles in certain regions of phase space, the resulting proper element distribution of inclinations will not naturally be fit by a von Mises-Fisher distribution, despite having likely originated as such.
This is demonstrated in Figure-\ref{fig:comp_prop}, which like Figure-\ref{fig:comp_fits}, shows a sample of particles selected according to the best-fit model of osculating inclination, the resulting stable population of particles in osculating elements space, and the resulting proper elements of that sample, as compared to the observed TNO population. 
Also shown in the third panel of Figure-\ref{fig:comp_prop} is the best-fit model for a direct fit of the mixed von Mises-Fisher model to the proper inclination distribution itself, rather than sampling some initial von Mises-Fisher population with our transfer function method.

When we compare the direct fit of the proper inclination distribution and compare against the best-model recovered from our transfer function, the two initially appear superficially very similar, but a fundamental difference becomes very noticeable.
We find that the direct fit to the proper inclinations requires a non-zero center, along with a modified $\kappa$ term which allows the width of the Cold Classical population to remain similar to the previous model.
In effect, the best-fit direct proper inclination model is near identical in shape to the transfer function model, but with a peak and upper wing which is shifted slightly upwards in inclination.
It can also be seen that, even in the case where we fit the proper inclinations directly with our mixed von Mises-Fisher distribution, the proper inclination of the observed CCTNO catalog does not seem to match the shape of a von Mises-Fisher distribution.
The reason why is naturally attributable to the secular instability caused by secular resonance, and secular forcing near the Laplace plane, which selectively lift particles to higher proper inclinations, and removing other particles entirely.
This systematic offset cab be somewhat captured by a static von Mises-Fisher distribution by way of shifting the center of the distribution upwards, but it is apparently evident that the additional feature of a flatter proper inclination distribution near $1.5^{\circ}<I_{free}<2.5^{\circ}$ remains.

Instead, when we sample the initial population in osculating element space with a smooth von Mises-Fisher distribution, the resulting proper inclination distribution is a much closer match to the observed TNO proper elements, since the present population has been shaped by these dynamical secular effects since the formation of the CCTNOs.

\begin{table*}[]
\begin{center}
    
\hspace{-1.5cm}

\renewcommand{\arraystretch}{1.6}
\begin{tabular}{|c|c|ccccc|cc|}
\hline
    $ $ & Sample Size & $\mu_{CC} (^\circ)$ & $\kappa_{CC}$ & $\mu_{HC}(^\circ)$ & $\kappa_{HC}$ & $x$ & $\sigma_{CC}(^{\circ})$& $\sigma_{HC}(^{\circ})$\\
\hline
\parbox[c]{3.2cm}{\centering Inner-main Belt  $(42.3 < a \,\scriptstyle{(au)}\textstyle< 43.7)$} & 347 & $0.507^{+0.39}_{-0.35}$ & $7.411^{+0.18}_{-0.12}$ & $9.300^{+6.84}_{-5.66}$ & $2.477^{+1.80}_{-1.67}$ & $0.837^{+0.03}_{-0.03}$ & $1.408^{+0.09}_{-0.12}$ & $16.623^{+21.83}_{-9.86}$\\
\hline
\parbox[c]{3.2cm}{\centering Kernel-main Belt  $(43.7 < a\,\scriptstyle{(au)}\textstyle< 44.5)$} & 347 & $0.782^{+0.36}_{-0.50}$& $7.576^{+0.28}_{-0.19}$ & $8.101^{+7.26}_{-4.93}$ & $2.784^{+1.99}_{-1.88}$ & $0.912^{+0.02}_{-0.02}$& $1.297^{+0.13}_{-0.17}$& $14.177^{+22.13}_{-8.92}$ \\
\hline
\parbox[c]{3.2cm}{\centering Outer-main Belt  $(44.5 < a\,\scriptstyle{(au)}\textstyle < 47.8)$} & 350 & $0.693^{+0.43}_{-0.47}$& $7.423^{+0.36}_{-0.27}$ & $4.092^{+2.10}_{-2.63}$ & $4.877^{+0.80}_{-0.68}$ &$0.492^{+0.07}_{-0.07}$& $1.401^{+0.20}_{-0.23}$& $4.092^{+2.10}_{-2.63}$ \\
\hline
\parbox[c]{3.2cm}{\centering Inner+Kernel Belt  $(42.3 < a\,\scriptstyle{(au)}\textstyle < 44.5)$} & 694 & $0.550^{+0.37}_{-0.37}$& $7.454^{+0.19}_{-0.11}$ & $8.791^{+6.97}_{-5.18}$ & $2.535^{+1.87}_{-1.71}$ &$0.877^{+0.02}_{-0.02}$& $1.379^{+0.08}_{-0.12}$& $16.161^{+21.88}_{-9.77}$ \\
\hline
\parbox[c]{3.1cm}{\centering All Classicals  $(42.3 < a\,\scriptstyle{(au)}\textstyle < 47.8)$} & 1044 & $0.484^{+0.37}_{-0.33}$& $7.372^{+0.17}_{-0.10}$ & $5.594^{+2.23}_{-3.48}$ & $4.308^{+1.21}_{-1.54}$ &$0.764^{+0.03}_{-0.03}$& $1.437^{+0.08}_{-0.12}$& $5.594^{+2.23}_{-3.48}$ \\

\hline

\end{tabular}

\end{center}
\caption{Median parameter posterior distribution fits to the 2-component models for the TNOs below $I<10^{\circ}$. We note that while an inclination center at $\mu_{CC}\approx0.5^{\circ}-0.8^{\circ}$ is the resulting median for every fit of the Cold Classical belt population, a center on the invariable plane, or $0^{\circ}$, is within $2\sigma$ in every case, with the best-fit parameters all lying near $0^\circ$. The Hot component of the Outer-main belt fit is the only region to be more tightly constrained, due to a denser population of higher inclination particles in that region. The last two columns show the corresponding median $\sigma$ widths generated from the model posterior distribution by way of $\kappa=1/\sigma^2$. Though this is not a valid transformation given a von Mises-Fisher value for $\mu\neq0^{\circ}$, the generated $\sigma$ values still provide reasonable insight into the actual width of the best-fit components, and are more intuitive.}
\end{table*}

\subsection{Comparison to Most Recent Results}
It's clear in the previous section that at $\mu_{CC} \approx0^{\circ}$ for each model, a Cold Classical belt width of $I \approx 1.5^{\circ}$ is generally preferred.
As discussed in Section~\ref{sec:fitting_results}, this differs from the width published by \cite{VanLaerhoven:2019} using the OSSOS dataset of $\sigma_{CC}\approx 1.75^{\circ}$, with the primary reasons relating to the sample size of the OSSOS dataset and the difference between the method implemented by each study.

To make a more direct comparison between the OSSOS result and our own result, we perform two separate analyses: (1) We measure the width of the OSSOS subset using our new method implementing the transfer function described in this paper, and (2) We directly fit the proper inclination distributions of both the OSSOS subset and our full dataset with a 2-component model, which fixes the inclination centers in the mixed von Mises-Fisher distribution to $\mu=0^{\circ}$, effectively fitting a mixed 2-component model of Rayleigh distributions.
We will correspondingly refer to these separate model fit analyses as Model-1 and Model-2, with respect to the OSSOS and Full datasets.

For brevity, we will only perform these comparisons for a single semi-major axis range.
We select the Inner-main + Kernel belt range of $42.3 < a < 44.5$ au, as it is the most densely populated region of the Classical belt, and is also the region most strongly impacted by the presence of the $\nu_8+\nu_{18}$ secular resonance, which primarily sculpts the proper inclination distribution over secular timescales.

We collect our subsample of OSSOS TNOs (which encompasses discoveries from \citealt{Petit:2011,Bannister:2016,Bannister:2018,Petit:2017,Alexandersen:2016}) contained in our data by retrieving the MPC designations contained in the Correspondence.list file published at \url{https://www.canfar.net/storage/vault/list/OSSOS/0_OSSOSreleases/OSSOSv12/ObsSummary}, and cross-referencing all of the designations with TNOs in our proper elements catalog.
A small number of the object in this set are actually defined as Centaurs in the MPC classifications and are not included in our catalog TNO proper elements. 
Of the 1124 tracked objects in this file, we find 1089 TNOs which are contained in our catalog, with the remaining objects consisting entirely of Centaurs included in the Correspondence.list file, which are not contained in our catalog.
This sample allows us to effectively sample nearly the entire OSSOS dataset for direct comparison to the \cite{VanLaerhoven:2019} analysis.
Within this subset, we find 256 OSSOS TNOs which lie within the Inner-main + Kernel Belt region with inclinations $I_{free} \leq 10^{\circ}$ for our comparison.

\begin{figure*}
    \centering
    \includegraphics[width=1\linewidth]{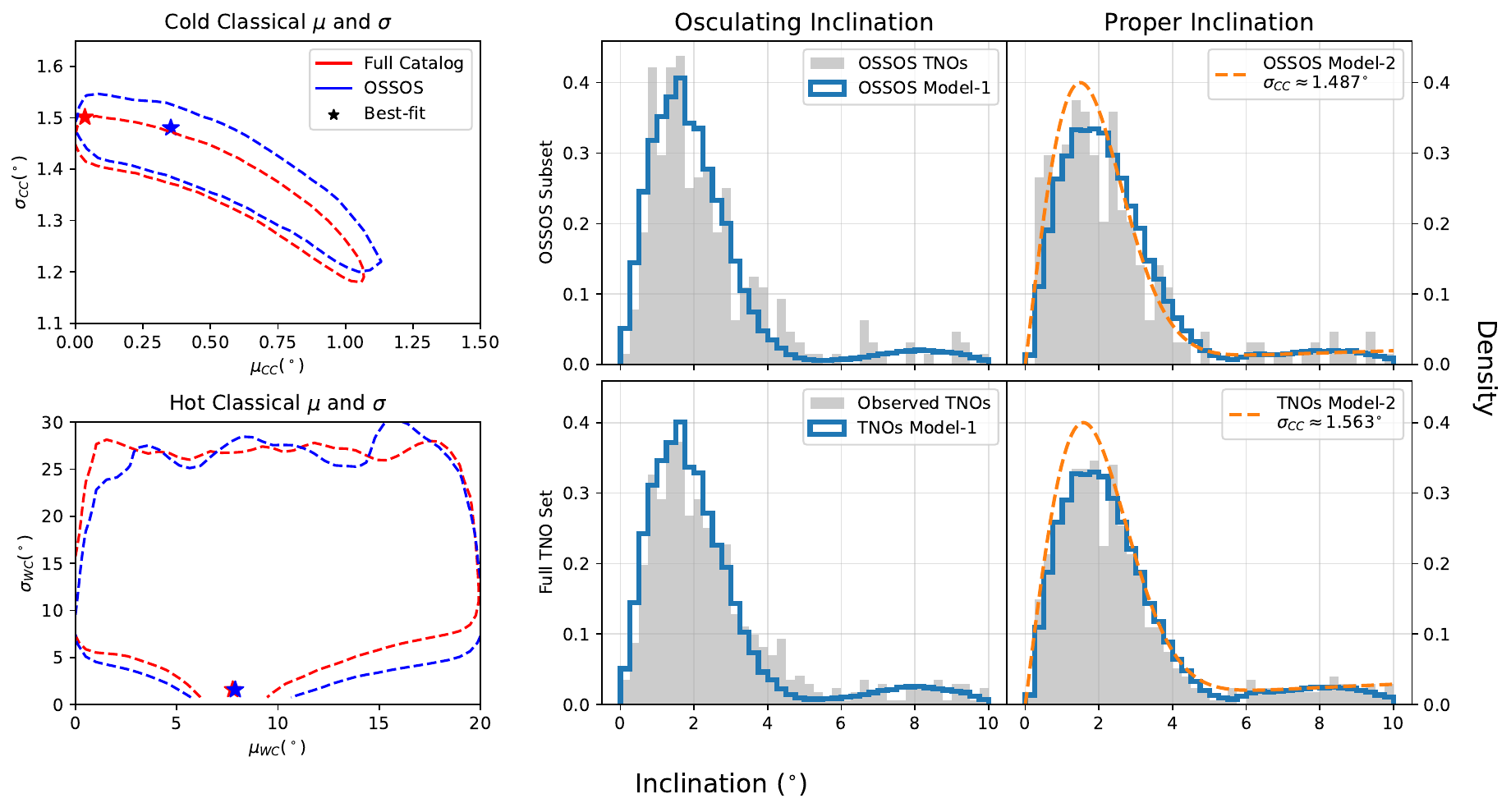}
    \caption{Comparison of best-fit models to the sampled OSSOS dataset and the full TNO dataset for the $42.3<a<44.5$ au, $I_{free}<10^{\circ}$ region of the TNO Classical belt. Model-1 represents the best-fit models for the transfer function method we implement in this paper. Model-2 represents a direct fit of our mixed vMF model to the proper inclinations of the corresponding datasets, which fixes the centers of the components at $\mu=0^{\circ}$, and as such is only shown in the proper element space. The left panels show the $1\sigma$ contours for the Model-1 posterior distribution of parameters for the OSSOS and Full Catalog datasets. The OSSOS dataset is very slightly better-fit by wider Cold Classical components than the full dataset in Model-1, though the bestfit Model-2 is actually slightly thinner than for the full catalog. Discrepancy between the \cite{VanLaerhoven:2019} result and our result is likely due to differences in the method, as well potentially noise among the smaller OSSOS sample. }
    \label{fig:ossos_comp}
\end{figure*}

The results of both comparisons are shown in Figure~\ref{fig:ossos_comp}, which displays the 1-sigma boundaries of the Model-1 fits, and the corresponding distributions of TNO inclinations in osculating and proper inclination space on the right for both the Model-1 and Model-2 fits.

With respect to the Model-1 fits, we can see in the top-left panel of Figure~\ref{fig:ossos_comp} that the resulting best-fit contour for the OSSOS dataset lies on average around $0.15^{\circ}$ higher than the same contour produced by our Full catalog sample.
The Model-2 fits have the same behavior, with a best-fit OSSOS Model-2 width of $\sigma_{CC}\approx1.487^{\circ}$, with a slightly thinner Full dataset fit of  $\sigma_{CC}\approx1.563$, which is a less severe difference in model fits.

While small, in both cases, the OSSOS model favors a wider Cold Classical belt component.
It is even visually noticeable that the OSSOS dataset includes a larger presence of particles near $I_{free}=4^{\circ}$, which may result from some bias introduced by the limited number of on-sky observing block location in nodal parameter space that contribute to the OSSOS cold classical observations.

While present in the Full catalog sample as well among the osculating inclinations, the contribution of this collection of objects appears to contribute to a more significant collection of $I_{free}>2^{\circ}$ TNOs than is seen within the Full catalog sample, effectively causing this sample to appear wider in it's proper inclination distribution.

In either case, as mentioned in Section~\ref{sec:fitting_results}, the results of the two studies remain quite close in comparison, and overlap with reasonable range of the reported uncertainties.
However, our method of computing the transfer function from a smooth von Mises-Fisher distribution of TNOs into the proper inclination space after 4.5 Gyr of evolution is recommended for future analyses of the belt, as the resulting proper inclination distribution for TNOs within the Cold Classical belt is not generally representative of a smooth von Mises-Fisher distribution, due to secular instability and migration, which our presented method accounts for.

\end{appendices}

\end{document}